\documentclass[cernpreprint, atlasdraft=false, USenglish, texmf, orcidlogo]{atlasdoc}
\usepackage[subcaption]{atlaspackage}
\usepackage{atlasbiblatex}

\usepackage{atlasphysics}
\usepackage{comment}

\graphicspath{{logos/}{figures/}}

\usepackage{ANA-HMBS-2024-34-PAPER-defs}
\AtBeginDocument{\RenewCommandCopy\qty\SI}

\AtlasTitle{Weakly supervised anomaly detection for resonant new physics in the dijet final state using proton--proton collisions at $\sqrt{s}=13$~TeV with the ATLAS detector}

\AtlasAbstract{%
An anomaly detection search for narrow-width resonances beyond the Standard Model that decay into a pair of jets is presented.
The search is based on 139 fb$^{-1}$ of proton--proton collisions at $\sqrt{s}=13$~TeV recorded during 2015--2018 with the ATLAS detector at the Large Hadron Collider.
The analysis is optimized without a particular signal model and aims to be sensitive to a broad range of new physics.
It uses two different machine learning strategies to estimate the background in different signal regions.
In each region, a weakly supervised classifier is trained to distinguish this background model from data.
The analysis focuses on events with high transverse momentum jets reconstructed as large-radius jets.
The mass and substructure of these jets are used as inputs to the classifiers.
After a classifier-based selection, the distribution of the invariant mass of the two jets is used to search for potential local excesses.
The model-independent results of both the anomaly detection methods show no signs of significant local excesses.
In addition to model-independent results, a representative set of signal models is injected into the data, and the sensitivity of the methods to these scenarios is reported.
}

\AtlasRefCode{HMBS-2024-34}

\PreprintIdNumber{CERN-EP-2025-002}

\AtlasJournal{Phys. Review D}
\AtlasJournalRef{\PRD 112 (2025) 072009}
\AtlasDOI{10.1103/2yq5-vj59}

\hypersetup{pdftitle={ATLAS document},pdfauthor={The ATLAS Collaboration}}

\begin{document}

\maketitle

\tableofcontents

\section{Introduction}

From the existence of dark matter to the matter-antimatter asymmetry, there are compelling motivations to search for particles beyond the Standard Model (BSM).
However, there are too many possibilities to search for all models individually.
Anomaly detection (AD) has emerged as an alternative approach to dedicated searches, capable of broad sensitivity to complement the deep sensitivity of targeted approaches.
Machine learning has catalyzed the development of new methods, with the ability to holistically explore high-dimensional spaces~\cite{Kasieczka:2021xcg,Aarrestad:2021oeb,Karagiorgi:2022qnh}.
Two broad classes of AD methods have emerged: \textit{weakly supervised AD} which uses weakly supervised machine learning to distinguish a target data sample from a reference data sample~\cite{HDBS-2018-59}; and \textit{unsupervised AD}  which uses unsupervised machine learning to identify events that are rare~\cite{Kasieczka:2022naq}.
This study explores the former, searching for a narrow resonance in dijet events from proton--proton ($pp$) collisions at $\sqrt{s}=13$~\TeV collected by the ATLAS detector at the Large Hadron Collider (LHC), which lasted from 2015 to 2018.
A weakly supervised approach, implemented in the Classification Without Labels (\Cwola) paradigm~\cite{Metodiev:2017vrx}, was chosen as it provides sensitivity to anomalous overdensities in the data, even when individual events cannot be identified as anomalous.
The analysis is model-agnostic and its primary goal is to highlight regions that might contain a potential BSM signal.

Resonant dijet events are a natural final state for new particle searches because many BSM scenarios feature massive particles that produce a pair of jets~\cite{Pappadopulo_2014,Craig:2016rqv,Kim_2020}.
These events are also well-suited for weakly supervised AD because data samples with little to no signal contamination can be constructed and used to build the reference data sample.
A previous ATLAS search in this final state~\cite{HDBS-2018-59} used weakly supervised AD built on the proposal of Refs.~\cite{Collins:2018epr,Collins:2019jip} to use the off-resonance sideband of the dijet invariant mass, or simple sideband, as the reference data sample directly.
Using the masses of the two jets to train the classifier, competitive limits were set on some signal models where a massive vector boson decays into two other vector bosons; these each decay into pairs of quarks, producing jets with two-prong substructure.
The main drawback of this approach is that it requires the classifier features to be nearly independent of the dijet mass.
This limits the number of features that can be used by the classifier and thus reduces the breadth of the search.

Several proposals address this challenge.
One phenomenologically well-studied technique uses the dijet mass sideband information for the interpolation into the signal region when forming the reference data sample~\cite{Nachman:2020lpy,Andreassen:2020nkr,Stein:2020rou,Benkendorfer:2020gek,Hallin:2021wme,Hallin:2022eoq,Chen:2022suv,Raine:2022hht,Golling:2022nkl,Sengupta:2023xqy,Das:2023bcj,Golling:2023yjq}.
This technique goes beyond the previous search by correcting for differences between the background in the signal region and sideband region that vary smoothly with the dijet mass.
The present analysis considers two such strategies: Simulation-Assisted Likelihood-free Anomaly Detection (\Salad)~\cite{Andreassen:2020nkr} and Constructing Unobserved Regions by Transforming Adjacent Intervals Flows for Flows (\Curtains)~\cite{Sengupta:2023xqy}.
\Salad uses a dijet-mass conditional reweighting function trained in the dijet mass sidebands to correct simulation in the signal region.
The reweighted simulation forms the reference data sample.
\Curtains uses a dijet-mass conditional morphing function trained in the dijet mass sidebands to correct the data in the sidebands to look like the data in the signal region.
The transformed sideband data forms the reference data sample.
Both the approaches correct for correlations between the dijet mass and the classifier features and thus can accommodate more features than were possible in the previous round of the analysis.
Accordingly,  the substructure features of the two jets in addition to the jet masses  are used   to train the classifier.

Other classical and AD searches have also probed hadronic final states.
Inclusive dijet searches~\cite{EXOT-2015-02,CMS-EXO-15-001} are broadly sensitive while dedicated searches for certain final states have significantly deeper sensitivity to certain types of models, including diboson resonances~\cite{HDBS-2018-11,HDBS-2018-19,HDBS-2018-31,HDBS-2020-19}.
The CMS Collaboration has performed three-dimensional bump hunts in the dijet and single jet mass spectra, targeting diboson-like decays~\cite{CMS-B2G-18-002,CMS-B2G-20-009}, and a model-agnostic search for dijet resonances with anomalous jet substructure~\cite{CMS-EXO-22-026}.
The ATLAS Collaboration has performed unsupervised AD searches for events with anomalous jets~\cite{HDBS-2019-23} and other resonant activity~\cite{EXOT-2022-07}.
The unsupervised AD approaches are complementary to the weakly supervised approach~\cite{Collins:2021nxn}.

The paper is organized as follows.
\Cref{sec:detector} introduces the ATLAS detector, \cref{sec:data} describes the data and simulated event samples used for the search, and \cref{sec:reconstruction} describes the event reconstruction and selection.
The analysis strategy is described in \cref{sec:analysis} and results are presented in \cref{sec:result}.
In addition to using new methods that enable more classifier features, this round of the search also expands on the interpretation of the result in two ways: more signal models are used to probe the sensitive region of BSM models and discovery reach in addition to exclusion is presented.
The paper ends with conclusions and outlook in \cref{sec:conclusion}.

\section{ATLAS detector}\label{sec:detector}
The ATLAS detector~\cite{PERF-2007-01} at the LHC covers nearly the entire solid angle around the collision point.\footnote{\AtlasCoordFootnote}
It consists of an inner tracking detector surrounded by a thin superconducting solenoid, electromagnetic and hadronic calorimeters,
and a muon spectrometer incorporating three large superconducting air-core toroidal magnets.

The inner-detector system (ID) is immersed in a \qty{2}{\tesla} axial magnetic field
and provides charged-particle tracking in the range \(|\eta| < 2.5\).
The high-granularity silicon pixel detector covers the vertex region and typically provides four measurements per track,
the first hit generally being in the insertable B-layer (IBL) installed before Run~2~\cite{ATLAS-TDR-19,PIX-2018-001}.
It is followed by the SemiConductor Tracker (SCT), which usually provides eight measurements per track.
These silicon detectors are complemented by the transition radiation tracker (TRT),
which enables radially extended track reconstruction up to \(|\eta| = 2.0\).
The TRT also provides electron identification information
based on the fraction of hits (typically 30 in total) above a higher energy-deposit threshold corresponding to transition radiation.

The calorimeter system covers the pseudorapidity range \(|\eta| < 4.9\).
Within the region \(|\eta|< 3.2\), electromagnetic calorimetry is provided by barrel and
endcap high-granularity lead/liquid-argon (LAr) calorimeters,
with an additional thin LAr presampler covering \(|\eta| < 1.8\)
to correct for energy loss in material upstream of the calorimeters.
Hadronic calorimetry is provided by the steel/scintillator-tile calorimeter,
segmented into three barrel structures within \(|\eta| < 1.7\), and two copper/LAr hadronic endcap calorimeters.
The solid angle coverage is completed with forward copper/LAr and tungsten/LAr calorimeter modules
optimised for electromagnetic and hadronic energy measurements respectively.

The muon spectrometer (MS) comprises separate trigger and
high-precision tracking chambers measuring the deflection of muons in a magnetic field generated by the superconducting air-core toroidal magnets.
The field integral of the toroids ranges between \num{2.0} and \qty{6.0}{\tesla\metre}
across most of the detector.
Three layers of precision chambers, each consisting of layers of monitored drift tubes, cover the region \(|\eta| < 2.7\),
complemented by cathode-strip chambers in the forward region, where the background is highest.
The muon trigger system covers the range \(|\eta| < 2.4\) with resistive-plate chambers in the barrel, and thin-gap chambers in the endcap regions.

The luminosity is measured mainly by the LUCID--2~\cite{LUCID2} detector that records Cherenkov light produced in the quartz windows of photomultipliers located close to the beampipe.

Events are selected by the first-level trigger system implemented in custom hardware,
followed by selections made by algorithms implemented in software in the high-level trigger~\cite{TRIG-2016-01}.
The first-level trigger accepts events from the \qty{40}{\MHz} bunch crossings at a rate close to \qty{100}{\kHz},
which the high-level trigger further reduces in order to record complete events to disk at about \qty{1.25}{\kHz}.

A software suite~\cite{SOFT-2022-02} is used in data simulation, in the reconstruction
and analysis of real and simulated data, in detector operations, and in the trigger and data acquisition
systems of the experiment.

\section{Data and simulated event samples}\label{sec:data}
The analysis uses data from $pp$ collisions at $\sqrt{s}=\qty{13}{\TeV}$ recorded with the ATLAS detector at the LHC during 2015--2018.
The total integrated luminosity of this data sample is \qty{139}{\per\femto\barn}~\cite{DAPR-2021-01} after data quality requirements~\cite{DAPR-2018-01}.
Events are recorded using the unprescaled single large-radius jet triggers with the lowest $p_\mathrm{T}$-threshold~\cite{TRIG-2016-01}.
For the trigger selection to be fully efficient, events are required to have at least one jet with a transverse momentum of $p_T>500$~\GeV, and they are required to satisfy $\mJJ > 1.3$~\TeV, where the invariant mass \mJJ is calculated according to the definition in \cref{sec:reconstruction}.

Monte Carlo (MC) simulations~\cite{SOFT-2010-01} are used to: (1) generate a background model using \Salad, (2) validate the analysis procedure, and (3) set representative model-dependent limits.
\PYTHIA[8.2]~\cite{Sjostrand:2014zea} is used as the nominal MC generator for the event samples.
Events are generated at leading order (LO) in QCD using the A14 tune~\cite{ATL-PHYS-PUB-2014-021} and NNPDF 2.3 LO~\cite{Ball:2012cx} parton distribution function (PDF) set.
The afterburner generator EvtGen~\cite{Lange:2001uf} is used to model decays of heavy flavor hadrons.
Detector effects are simulated using \GEANT~\cite{Agostinelli:2002hh}.
All simulated events are overlaid with additional minimum-bias events generated with \PYTHIA[8.186]~\cite{Sjostrand:2007gs} using the A3 tune~\cite{ATL-PHYS-PUB-2016-017} and the NNPDF 2.3 LO~\cite{Ball:2012cx} PDF set to simulate the effect of multiple $pp$ collisions per bunch crossing (pileup).
The distribution of the average number of pileup interactions in simulation is re-weighted during data analysis to match that observed in the Run 2 data.

Samples of MC simulated $2\rightarrow 2$ dijet events are used to emulate the main Standard Model (SM) background.
As the jet production cross-section is $\mathcal{O}(100)$ times greater than electroweak processes, the consideration of these samples is sufficient to describe the data and all other processes are negligible.
To fully populate a wide range of jet \pt, these samples are generated in slices of the particle-level $R=0.6$ jet \pt~\cite{ATL-PHYS-PUB-2019-017}.

Samples of MC simulated events are used to emulate twenty different signal models.
The samples contain the processes $W'\rightarrow W''Z''$ and $A^0\rightarrow H''Z''$, for new vector bosons $W'$~\cite{Altarelli:1989ff} and pseudoscalar boson $A^0$~\cite{Branco:2011iw}.
Following Ref.~\cite{HDBS-2018-59}, the masses of the $W'$ and $A^0$ are considered to be either 3~\TeV or 4.5~\TeV, labeled as low-mass and high-mass respectively.

The $W''$, $Z''$, and $H''$ particles are altered versions of the SM $W$, $Z$, and Higgs bosons.
For the $W'\rightarrow W''Z''$ process, following Ref.~\cite{HDBS-2018-59}, the masses of the $W''$ and $Z''$ bosons are varied to be one of the specific values: 80~\GeV, 200~\GeV, and 400~\GeV.
For the $A^0\rightarrow H''Z''$ process, the mass of the $H''$ boson is set to 200~GeV and the mass of the $Z''$ boson is set to 400~\GeV.
The widths of all altered $W''$, $Z''$, and $H''$ particles are set to \qty{0.1}{\GeV}.

In one set of signal samples, as in Ref.~\cite{HDBS-2018-59}, the $W''$ and $Z''$ bosons decay hadronically with $W''/Z'' \to q\bar{q}$ for non-top quarks $q$ and $\bar{q}$, and same-flavor $q/\bar{q}$ for the $Z''$ decays.
An additional set of signals includes $H''\rightarrow b\bar{b}$ or $H''\rightarrow \gamma\gamma$ and $W''/Z''\rightarrow qqq$ decays.

The $W' \to W''Z''$ sample with $m_{W'} = 3$~\TeV, $m_{W''} = 200$~\GeV, and $m_{Z''} = 400$~\GeV is simulated twice, with different $\mJJ$ spectra arising from different parton shower settings.
This is done to explore the effect of the $\mJJ$ spectrum on signal sensitivity.
As is shown in \cref{sec:exclusion}, the analysis is more sensitive to the narrower resonance.

Additionally, one set of signal models is simulated from the process $V' \to WV$, with $V'$ a heavy vector boson from a Heavy Vector Triplet model~\cite{Pappadopulo_2014} and $V=W,Z$.
The mass of the parent particle ($m_{V'}$) is varied to be one of the specific values 2.6~\TeV, 2.8~\TeV, and 3.0~\TeV, and the daughter particles decay hadronically.

The signal models are named $P^{s,d}_{p,a,b}$, where $P$ is the name of the parent particle and  the mass of the parent particle in \GeV is denoted by $p$\@.
If applicable, the masses of the daughter particles in \GeV are denoted by the subscripts $a,b$.
If applicable, whether the signal spectrum is narrow ($n$) or wide ($w$) is denoted by the superscript $s$.
And, also if applicable, the decay channel is denoted by the superscript $d$.
For example, $W^{\prime n}_{3000,80,400}$ denotes a 3~\TeV $W'$, which decays into a 80~\GeV $W''$ and a 400~\GeV $Z''$, with a narrow $\mJJ$ spectrum.

\section{Event reconstruction and selection}
\label{sec:reconstruction}

Events are required to contain at least one reconstructed $pp$ collision vertex candidate with at least two associated ID tracks with \pt larger than 0.5 GeV \cite{ATL-PHYS-PUB-2015-026}.
The primary vertex (PV) for each event is chosen as the reconstructed vertex with the highest sum of the $\pt^2$ of its associated tracks.
Jets are reconstructed from locally calibrated calorimeter cell-clusters~\cite{PERF-2014-07} using the anti-$k_t$ algorithm~\cite{Cacciari:2008gp,Fastjet} with a radius parameter of $R=1$.
These large radius jets are trimmed~\cite{Krohn:2009th} by reclustering the jet constituents with the $k_t$ algorithm using $R = 0.2$ and removing the constituents with \pt less than 5\% of the original jet \pt.
The jet four-vectors are then calibrated as detailed in Ref.~\cite{JETM-2018-02}.
The analysis uses a subset of the available kinematic features of the events, which is described in the following.
The dijet mass (\mJJ) is defined in this study as
$
\mJJ^2 = 2(|\vec{p}_1||\vec{p}_2| - \vec{p}_1 \cdot \vec{p}_2),
$
where the momenta are those of the two jets with the highest \pt.\footnote{This differs from the typical definition, $\mJJ^2 = (p_1^\mu + p_2^\mu)^2,$ by setting the individual jet masses to zero. The choice was made to reduce the correlation between \mJJ and the training features.}
The feature $M$ is defined as the mass of a single jet.
The feature $\tau_{21} = \tau_2 / \tau_1$ ($\tau_{32} = \tau_3 / \tau_2$) is the ratio of the 2- and 1-subjettiness (3- and 2-subjettiness)~\cite{Thaler:2010tr} of each jet, with lower values indicating that the jets are more two- (three-) prong.
The features $\tau_{21}$ and $\tau_{32}$ are selected from a larger set of jet substructure observables for their classification performance across the signal models described in \cref{sec:data}.
Including a specific jet feature in the analysis means incorporating its value for both the  jets.

The requirements used to select events are summarized in \cref{tab:event_selection}.
The leading jet $p_\mathrm{T}$ selection is based on the trigger, as described in \cref{sec:data}.
A requirement on the $p_\mathrm{T}$ of the subleading jet enhances events with a dijet topology.
The jets are required to be relatively central without extreme values of jet mass so that their substructures are well reconstructed.
The $s$-channel resonance signals which generate dijet events are targeted, as two jets is the primary decay channel for many new physics processes.
Only events where the absolute difference between the rapidity $y$ of the two jets, $\dY = |y_1-y_2|$, is less than 1.2 are considered, as this selection contains a higher proportion of $s$-channel events.
\begin{table}[ht]
\begin{center}
\caption{Summary of the event selection used to define the data sample for the analysis based on reconstructed jet properties.}
\label{tab:event_selection}
\begin{tabular}{l l}
\hline
Observable & Selection \\
\hline
leading jet $p_\mathrm{T}$ (\GeV) & $>500$\\
subleading jet $p_\mathrm{T}$ (\GeV) & $>200$\\
jet $|\eta|$ (both) & $<2.0$\\
jet mass, $M$ (\GeV) (both) & $30 < M < 500$\\
$\dY = |y_1-y_2|$ & $<1.2$\\
\hline
\end{tabular}
\end{center}
\end{table}

\section{Analysis method}\label{sec:analysis}
The analysis employs weakly supervised machine learning to search for resonant signals, which appear as localized peaks in the \mJJ spectrum.
The primary background arises from non-resonant dijet production via the strong interaction.
In addition to being resonant, a potential signal is expected to show distinct properties in other observables, like jet substructure, compared with the background, enabling effective differentiation.

\begin{figure}[tb]
\centering
\begin{subfigure}[t]{0.28\textwidth}
\centering
\includegraphics[width=\textwidth]{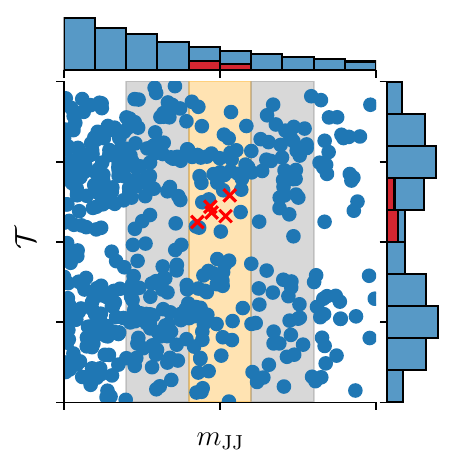}
\caption{Definition of Regions}
\end{subfigure}%
\begin{subfigure}[t]{0.28\textwidth}
\centering
\includegraphics[width=\textwidth]{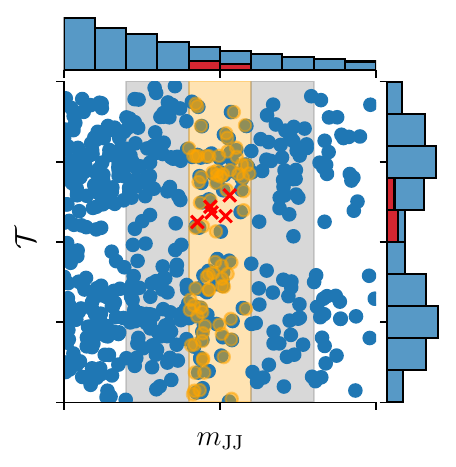}
\caption{Background Estimation}
\end{subfigure}%
\begin{subfigure}[t]{0.28\textwidth}
\centering
\includegraphics[width=\textwidth]{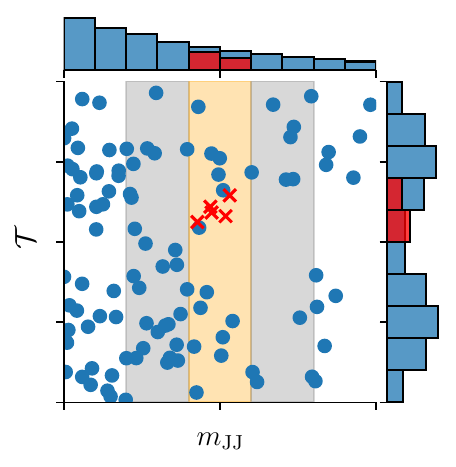}
\caption{Classification of Signal}
\end{subfigure}%
\begin{subfigure}[t]{0.16\textwidth}
\centering
\raisebox{1cm}{\includegraphics[width=\textwidth]{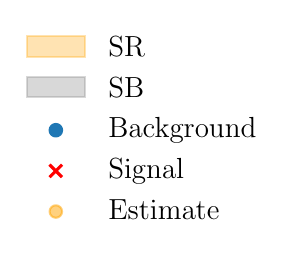}}
\end{subfigure}
\caption{
A schematic drawing of the analysis strategy with the consecutive steps shown in the different panels.
The distribution shows the data which consists of a background and a potential signal which is localized on the \mJJ spectrum and (potentially) different from the background in other features, here collectively labeled as $\mathcal{T}$.
The first panel (a) shows how the SR and SB regions are defined on the \mJJ spectrum.
The second panel (b) shows how the background is estimated in the SR\@.
Then a classifier is trained between the estimated background and the data in the SR\@.
The third panel (c) shows the data after a cut is applied on the classifier output of each event, thereby increasing the relative contribution of a potential signal.
After this step, a bump hunt is performed in the \mJJ spectrum.
The whole process is repeated with shifted regions SR and SB until the entire \mJJ spectrum is covered.
}
\label{fig:strategy_scheme}
\end{figure}

The analysis strategy is structured into four steps, whereas the first three steps are explicitly depicted in \Cref{fig:strategy_scheme}:
\begin{enumerate}
\item Potentially signal-enriched (Signal Region, SR) and signal-depleted regions (Sideband, SB) are defined.
\item Two different methods --- \Salad and \Curtains --- use the signal-depleted regions to create estimates of the background process in the signal-enriched region.
\item A weakly supervised machine learning classifier is trained to classify the estimated background from data, and enhances the ratio of a potential signal to the background
\item A statistical inference is performed in the signal-enriched region.
\end{enumerate}
Each step is briefly described in this overview, with more details in the following sections.

The analysis starts with the definition of the SR and SB\@.
As potential signals are localized in the \mJJ spectrum, SR and SB are defined as collections of events with \mJJ values in ranges on this spectrum.
Various mass hypotheses are systematically tested by iteratively shifting the SR and SB over a total of seven combinations of regions, which cover the investigated \mJJ spectrum.
For each combination of regions the entire analysis chain is independently executed, including the background estimations, the training and application of the weakly supervised classifier, and the statistical analysis, is employed separately.
A more detailed explanation of these definitions and the shifting procedure is provided in \cref{sec:regions}.

Different subsets of the event features introduced in \cref{sec:reconstruction} are used as the outputs of the background estimation and the inputs of the classification.
Three subsets $\mathcal{T}$ are used: $\{M\} \text{, } \{M, \tau_{21}\} \text{, and } \{M, \tau_{21}, \tau_{32}\}$.
Jet mass $M$ is chosen because it is known to be sensitive in anomaly searches~\cite{HDBS-2018-59}; $\tau_{21}$ and $\tau_{32}$ are included to increase sensitivity to boosted 2- and 3-body decays of BSM particles, reconstructed as single large-radius jets.
As the weakly supervised classifier (described below) uses these features to select events, a selection of features is akin to defining a signal region in a traditional analysis.
Adding more features may broaden the sensitivity to some signals, but it also makes the model training harder, and thus can weaken the sensitivity to particular signals.
The inclusion of uninformative features can generally also degrade the performance of a classifier as shown in Ref.~\cite{grinsztajn2022treebasedmodelsoutperformdeep}.
New strategies for classifier training that scale well with the number of features are being actively developed, as seen in Refs.~\cite{Finke:2023ltw,Freytsis:2023cjr,golling2024masked,Mikuni_2025}.

Because the SB is signal-depleted, it can be used to estimate the background-only distribution in the SR\@.
Two methods use neural networks to explicitly model the high-dimensional background distribution in the SR\@.
The features in $\mathcal{T}$ vary smoothly with \mJJ, and the validation methods show that the background distribution in the SR can be estimated from the SB\@.
This explicit modeling is an improvement to the previous weakly supervised analysis in ATLAS~\cite{HDBS-2018-59}, as it largely prevents an \mJJ dependent bias of the classification of a potential signal.
Both the methods are described in greater detail in \cref{sec:background}.

Next, a weakly supervised classification algorithm is trained to distinguish between the data distribution and the estimated background distribution in the SR\@.
Following the \Cwola paradigm~\cite{Metodiev:2017vrx}, this algorithm, implemented as a neural network, approximates an optimal classifier between a potential signal and the background.
A cut on the classifier output is then applied to enhance the fraction of signal events in the SR\@.
An ensemble of multiple classifiers ensures more stable results and mitigates the risk of false positives in the absence of a signal.
The classification is described in greater detail in~\cref{sec:CWOLA}.

In the final step of the analysis chain, a parametric function is fitted to the \mJJ spectrum in the SB and interpolated into the SR\@.
The inference framework then compares this interpolated prediction with the observed number of data events in the SR\@.
Observed significances of the analyzed data compared with the estimated number of background events are calculated using a profile likelihood fit.
Upper limits on the expected number of signal events for various signal models are established at a $95\%$ confidence level (CL) using the $CL_S$ method~\cite{Read:2002hq}.
The inference strategy is described in greater detail in \cref{sec:inference}.

In data-driven weakly supervised AD analyses like this one, the validation of analysis performance is especially critical as the presence of an unknown potential signal changes the method (i.e., the classifier) itself.
The validation strategy is explained in greater detail in \cref{sec:validation}.

\subsection{Definition of signal and background regions}\label{sec:regions}
The potential signals are resonances and therefore localized in the \mJJ spectrum.
Events are considered parts of the SR if their \mJJ values fall inside a $\pm 300$~\GeV window around the targeted signal mass.
The SB consists of events with \mJJ values up to \qty{600}{\GeV} below or above the SR edges.
Multiple hypotheses are tested by sliding the set of SR and SB over the \mJJ spectrum.
The analysis investigates the \mJJ spectrum between \qty{2.6}{\TeV} and \qty{5.0}{\TeV} in steps of \qty{300}{\GeV}.
The SR width was chosen to encompass the experimental width of narrow resonances at \qty{3}{\TeV},\footnote{For example, for the signal model $W' \to W'' Z''$ with $m_{W''} = m_{Z''} = \qty{200}{\GeV}$, the standard deviation of \mJJ is \qty{400}{\GeV}.} and it was shown that the \qty{600}{\GeV} window also covers the width of the high-mass signal models considered.
The choice of \qty{300}{\GeV} steps ensures that all SR overlap and guarantees a seamless testing of the entire \mJJ range.

The tested SRs with values of \mJJ in units of \TeV are: [2.6, 3.2], [2.9, 3.5], [3.2, 3.8], [3.5, 4.1], [3.8, 4.4], [4.1, 4.7], [4.4, 5.0].
The number of events decreases with increasing \mJJ and the highest SR was chosen such that there are more than ten events in data for each bin in the high \mJJ SB of this SR at the tightest tested classifier selection explained below.
Only the regions which satisfy the validation discussed in \cref{sec:validation} are considered for the final results of the analysis.

\subsection{Background model}\label{sec:background}
Two different methods are used to generate background templates for training the signal-identification classifier: \Salad and \Curtains.
The techniques are independent and complementary --- \Salad uses a simulated prior, whereas \Curtains is entirely data-driven.
As a result, the level of agreement between the \Salad and \Curtains branches of the analysis provides a consistency check on the results.

In the discussion below, suppose that a data sample is provided with features $\mathcal{T} \cup \{\mJJ\}$, where $\mathcal{T}$ is some set of features.
The goal is to use the SB to generate an \mJJ-parameterised background template which approximates the features $\mathcal{T}$ well in the SR\@.
Then the features $\mathcal{T}$ can be used for training the signal-identification classifier.
The specific architectures of all methods are detailed in \Cref{sec:implementation_appendix}.
For the lowest (highest) SR case, a total of \num{659\,059} (\num{6381}) SB events are used to train the \Salad and \Curtains models, and \num{115\,300} (\num{1559}) events are in the SR\@.

The \Salad~\cite{Andreassen:2020nkr} method generates a background model based on a high-fidelity simulation using the following steps:
\begin{enumerate}
\item Assign every event in the simulated SB a label of 0, and assign every event in the data SB a label of 1.
\item Train a neural network $A$ to classify simulation and data in the SB using $\mathcal{T} \cup \{\mJJ\}$ as training features. The classifier $A$ is parameterized in \mJJ~\cite{Cranmer:2015bka,Baldi:2016fzo} to enable an interpolation into the SR\@.
\item For each simulated event $i$ in the SR, let $A(i)$ be the output of classifier $A$ when it sees the event. Assign event $i$ a weight $$w_i = \frac{A(i)}{1 - A(i)},$$ which approximates the probability density ratio of data to simulation. The final \Salad background estimation in the SR is the SR simulation with the weight $w_i$ applied to every event.
\end{enumerate}

The \Curtains method uses the SB data to fit a model and generates a background model using the following steps:
\begin{enumerate}
\item Train a normalizing flow $p_\theta$ with the network parameters $\theta$~\cite{tabak2013family} to estimate the conditional density of the features in the SBs $p(\mathcal{T}|\mJJ \in\{\mathrm{SB}\})$.
\item \label{step:ctransformer} Train another normalizing flow to map from $p_\theta(\mathcal{T}|\mJJ = m_1)$ to $p_\theta(\mathcal{T}|\mJJ = m_2)$ for all $m_1,m_2$ pairs in the joint SBs.
\item \label{step:mass_fit} Fit the mass distribution in the SBs using the same baseline function used in the previous search~\cite{HDBS-2018-59}, $p_1(1-z)^{p_2}z^{p_3}$, where $z$ is the mass divided by the center of mass energy and $p_i$ are free parameters. Interpolate into the SR and sample~\cite{zfit}.
\item Use the normalizing flow from Step~\labelcref{step:ctransformer} to map each SB sample to a mass value in the SR sampled using the fit from Step~\labelcref{step:mass_fit}.
\end{enumerate}
Long-tailed features in $\mathcal{T}$ are transformed using $f(x) = \log(\frac{p}{1-p})$, also known as a logit transform, before being fed into the normalizing flow, as this is shown to improve the modeling performance~\cite{Hallin:2021wme}.
The last step can be repeated multiple ($m$) times, and if there are $n$ events in the combined SBs then \Curtains will generate $m \times n$ events in the SR\@.
This contrasts with \Salad, which will only generate as many events as there are simulated events in the SR\@.
The oversampling is shown to increase signal sensitivity~\cite{Hallin:2021wme, Raine:2022hht, Sengupta:2023xqy}.
Given the seven SRs and three feature sets, a total of 21 \Curtains models and 21 \Salad models are trained.

\subsection{Classification of a potential signal}\label{sec:CWOLA}
One fundamental challenge in any weakly supervised application is that both the training and evaluation data are imprecisely labeled.
Using the \Cwola paradigm~\cite{Metodiev:2017vrx} allows training of a classifier even without precise labels.
All that is needed to train a binary classifier is two data samples with different frequencies of the two respective classes.
As the two example data samples only differ in their signal-to-background composition, the classifier will learn to identify this one difference by separating signal events from background events.
Further, a classifier trained this way will converge to an optimal signal vs. background classifier, so long as the initial assumptions about the data samples only differing in composition are correct.

\subsubsection{Classifier implementation}
In the previous ATLAS weakly supervised search~\cite{HDBS-2018-59}, the \Cwola approach was used to search for anomalous signals by discriminating between (1) an SR with a high suspected signal contamination and (2) the SBs with low expected contamination.
One major challenge in this approach was the need to ensure decorrelation between the observable used to separate SR and SB and the observables used to train the classifier; otherwise, the classifier could learn the defining distinction between SR and SB and introduce artificial bumps which could form a false excess as a result, a process also known as background sculpting~\cite{Hallin:2022eoq}.

This analysis improves on the pure \Cwola search by instead training the classifier between (1) the data in a candidate SR and (2) a background template in that region, provided by the approaches described in \cref{sec:background}.
Using a direct modeling of the background in that region reduces the need for decorrelation compared with previous approaches.
The template is assumed to be signal-depleted, while the data are signal-enriched.
Following this assumption, a classifier, which is trained to distinguish the data and the template, will therefore give a higher classifier score to signal-like events.
The classifier architecture is defined in \Cref{sec:implementation_appendix}.

A five-fold cross-validation strategy is applied; this ensures that classifiers are not evaluated on the same data they are trained on.
In this setting the \Cwola model is trained to discriminate between a subset of the SR data and a generated template, then evaluated on the unseen portion of the SR data.
Two working points are defined for the selection efficiency of the \Cwola classifier --- one where the top 10\% of events with the highest classifier score are retained $\epsilon=0.1$, and a tighter selection where the top 2\% of events with the highest classifier score are retained $\epsilon=0.02$.
The classifier working points are chosen to enable sensitivity to a wide range of signals while ensuring enough SB events in the high-\mJJ region for a viable fit (described in \cref{sec:fitting strategy}).
Some signals are more sensitive to tighter cuts, some to looser.
For a given signal, the optimal cut is a function of the cross-section and characteristics of the signal, and is not predictable \textit{a priori}.

\subsubsection{Ensemble methods}\label{sec:expected sensitivity ensemble methods}
An ensembling method is used to minimize the dependence on the (randomized) classifier-weight initialization.
Many classifiers are trained for every SR, and each classifier is initialized with a different random seed.
Without any signal, the classifiers tend to randomly pick regions of phase-space while with a  signal, most of classifiers tend to agree on one region of phase-space~\cite{Golling:2023yjq}.

A strategy is developed to combine the classifiers, developed in an idealized setting, where the \Cwola classifier is trained to discriminate between two data samples drawn from the same distribution.
This setting is chosen as it removes issues related to template mis-modeling.
Using an ensemble of $N$ classifiers, the best strategy is found to be: first make a selection on the classifier output of each of the $N$ classifiers, then fill $N$ histograms in the \mJJ distribution, one for each selection, and then average these $N$ histograms using the median.
The size of the ensemble in this search is set to $N=10$.

The averaged histogram is the input to the fit and subsequent statistical analysis.
The average bin count is Poisson distributed with an additional uncertainty given by the standard error on the estimate of the median.
This strategy can be viewed as a marginalization over randomly placed decision boundaries that removes the bias from the classifier initialization.
Ensembling directly based on the output of the classifier was shown to either decrease the sensitivity of the analysis or increase the rate of false positives.

\subsection{Inference framework}
\label{sec:inference}
In the final step of the analysis strategy, a parametric function is fit to the \mJJ spectrum in the SB and interpolated to the SR\@.
This interpolation and the recorded data in the SR are used to estimate the statistical significance of potential deviations and to set limits on benchmark models.
This section first outlines the fitting strategy and the interpolation, then, it introduces the relevant uncertainties, explains the statistical analysis, and introduces the post-hoc non-closure correction of the significance.

\subsubsection{Fitting strategy}
\label{sec:fitting strategy}
The production of dijets in the SM is dominated by QCD processes, and is well-described by a smoothly falling distribution in \mJJ.
For a given SR, after making a classifier selection as described in \cref{sec:CWOLA} and defining a histogram in bins of \mJJ, a fit is made to the SB bins.
Thirty bins with a width of $60$~\GeV evenly span both the  SBs and the SR\@.
For the fit, the SR is masked.
The bins in the SB are then fit with an iterative strategy until the $p$-value from the $\chi^2$ fit is greater than 5\%.
The same fit strategy is used as in the previous ATLAS weakly supervised search~\cite{HDBS-2018-59}.
Upon each iteration, the fit function increases in complexity as follows: (1) a three parameter fit $dn/dx=p_1(1-x)^{p_2}x^{-p_3}$, (2) a four parameter fit $dn/dx=p_1(1-x)^{p_2}x^{-p_3+p_4\log(x)}$~\cite{EXOT-2019-03}, (3) the UA2 fit function $dn/dx=p_1x^{p_2}e^{-p_3x + p_4x^2}$~\cite{Alitti:1990kw}, where $x=\mJJ/\sqrt{s}$ and $\{p_i\}_{i=1}^4$ are the fit parameters.
If the fit $\chi^2$ criterion is still not met after cycling through all possible fit functions, then the SB window is reduced by removing the SB-bin furthest from the SR, for each SB\@.
This is repeated until the fit $\chi^2$ criterion is met, or until the SB contains fewer than six bins, in which case the fit is considered to have failed.
This kind of fit failure did not occur in practice.

\subsubsection{Uncertainties}
\label{sec:uncertainties}
Both the observed data and the estimated number of background events, which are used in the statistical analysis, contribute a set of uncertainties.

The observed data contributes two different uncertainties: the Poisson statistical uncertainty in the bin count and the uncertainty from the classifier ensemble described in \cref{sec:expected sensitivity ensemble methods}.
The Poisson statistical uncertainty in the bin count is inferred as $\sqrt{n}$ with the bin count $n$ being the average over the classifier ensemble.
The uncertainty from the classifier ensemble is driven by the randomness that results from the different classifier initializations.
It is determined for each bin by the spread of the bin counts from the classifiers in the ensemble and includes the correlation between the classifiers.
The uncertainty from the classifier ensemble is propagated into the uncertainty on the bin count by summing it in quadrature with the Poisson statistical uncertainty.

The uncertainty in the estimated number of background events then comes from the fit to the \mJJ spectrum, which includes the uncertainties on the bin counts described above.
After the fit is performed, the fit uncertainty in the bins in the SR is determined by propagating the uncertainty in the fit parameters to the individual bins.

As highlighted above, the primary goal of this weakly supervised search is to look for hints of new physics, not to set limits on particular new physics models.
Additionally, the search probes a region of phase-space in which the signal contribution is mostly statistically limited.
Therefore, systematic uncertainties in the injected signals are not considered.

The largest uncertainty in all hypothesis tests is the uncertainty on the estimated number of background events, derived from the fit to the \mJJ spectrum.
The next largest contribution is the Poisson statistical uncertainty in the data which is slightly larger than the uncertainty from the ensembling of the \mJJ histograms of the data.

\subsubsection{Likelihood fit}
\label{sec:likelihood fit}
To extract a significance on the observed SR counts, a profile likelihood fit is used~\cite{Collins:2019jip,Cowan:2010js}.
The bins of the \mJJ histograms that lie in the SR are combined into a single bin.
All sources of uncertainties in the likelihood fit are described in \cref{sec:uncertainties}.
The uncertainty from the fit function and ensemble procedure are included as Gaussian distributed constrained uncertainties.
All likelihood fits are performed using \texttt{pyhf}~\cite{pyhf_joss_fix,pyhf_zenodo}.

\subsubsection{Non-closure correction}
\label{sec:nonclosure}
Two systematic effects can cause a non-closure of the analysis methodology.
The first effect comes from the uncertainty in the function used to fit the \mJJ spectrum.
The second effect comes from possible deficiencies in the template generation from \Salad or \Curtains.
Such deficiencies could, for example, introduce a systematic bias which could be identified by the classifier.
This bias could cause a false excess, which could be caught by the validation described in \cref{sec:validation}, but it is more likely to reduce the sensitivity of the search.

In each SR, the possible non-closure is corrected with a post-hoc non-closure correction on the extracted significance, effectively adding a post-evaluation uncertainty.
The correction uses the \downupdataset test data samples described in~\cref{app:testdatasetsvalidation:datasets}.
The statistical properties of the $p$-value distribution are assessed on these test data samples and derive a calibration procedure to correct for any biases of the analysis.
A linear fit to the mean and standard deviation of the significance across the ten different \downupdataset test data samples as a function of \mJJ is used to derive a correction to the final significance.
The fitted significance is subtracted from the significance as evaluated on the data in the SR\@.
The correction is only used to decrease the significance with the addition that it cannot (further) decrease the significance below zero.
The corrections are small everywhere, with no correction to the standard deviation.
The largest shift in the mean is of 0.5$\sigma$.

\subsubsection{Limit setting}
Upper limits on the production of various signal models are set at $95\%$ confidence level (CL) using the $CL_S$ method~\cite{Read:2002hq}.
For each signal model, limits are set by injecting a fixed number of signal events into the data and then running the full analysis, using the injected signal events to calculate the expected number of signal events after a selection is performed.
Multiple different signal injections are performed to build curves for the observed, expected, and $\pm 1\sigma$ and $\pm 2\sigma$ bands.
The intersection of each of these curves with the horizontal line at $95\%$ CL is used to set the limit.

\subsection{Validation}
\label{sec:validation}
Validating a data-driven, weakly supervised anomaly detection search is inherently challenging because the presence, magnitude, and characteristics of potential signals are unknown; however, the method's performance depends on the input data, which may include these potential signals.
The analysis uses three complementary and orthogonal test data samples for validation, referred to as \deltaydataset, \mcdataset, and \downupdataset respectively.
All three test data samples are described in greater detail in \cref{app:testdatasetsvalidation:datasets}.
Each of the test data samples is designed to test a different aspect of the analysis and the use of all three provides confidence in the robustness of the analysis.

Each \mJJ SR is validated individually using all three test data samples as explained in the following.
For each \mJJ SR, the significance of the respective data above the estimated number of background events is evaluated with each of the test data samples.
An \mJJ SR is considered to satisfy the validation for a test data sample if the median significance is less than one and the spread in the significances is less than one on this given test data sample.
An \mJJ SR has to satisfy the validation for each of the test data samples to satisfy the overall validation and to be considered for further analysis.

Seven of the original eight \mJJ SRs satisfy the validation and are considered for further analysis.
The lowest \mJJ SR does not satisfy the validation and is excluded from the final results.
This validation failure can be attributed to the correlations between \mJJ and the masses of the two jets, which are shown to differ between the low and high \mJJ regions.
These correlations make the background modeling more challenging in the low \mJJ region.
More details of the validation procedure and the individual results of the validation procedure are shown in \cref{app:testdatasetsvalidation:validation}.

\section{Results}
\label{sec:result}
\begin{figure}[tb]
\centering
\begin{subfigure}[b]{0.3\textwidth}
\centering
\includegraphics[width=\textwidth]{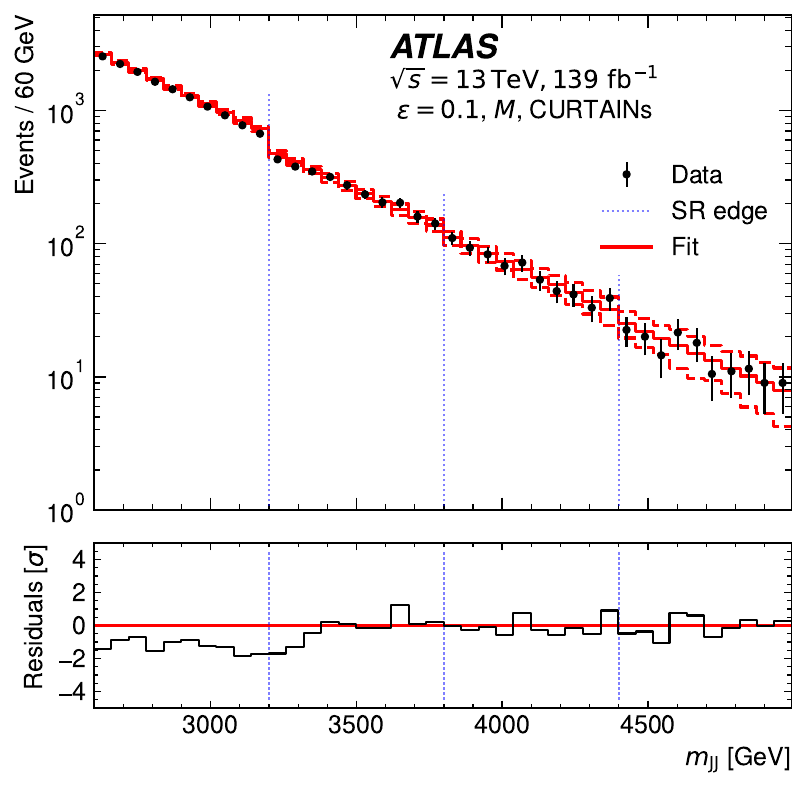}
\caption{}
\end{subfigure}
\hspace{0.01\textwidth} %
\begin{subfigure}[b]{0.3\textwidth}
\centering
\includegraphics[width=\textwidth]{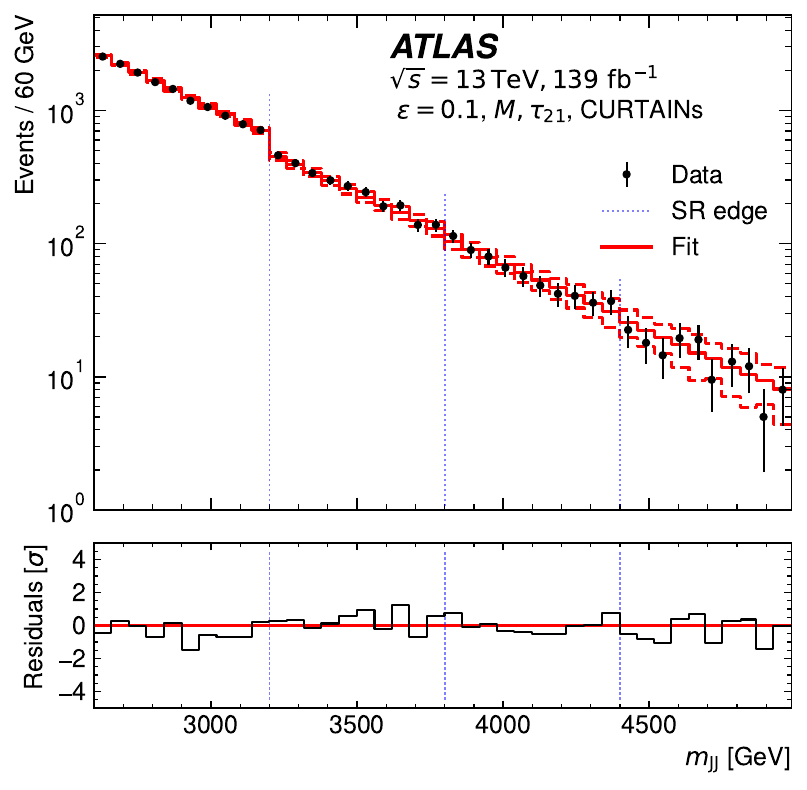}
\caption{}
\end{subfigure}
\hspace{0.01\textwidth} %
\begin{subfigure}[b]{0.3\textwidth}
\centering
\includegraphics[width=\textwidth]{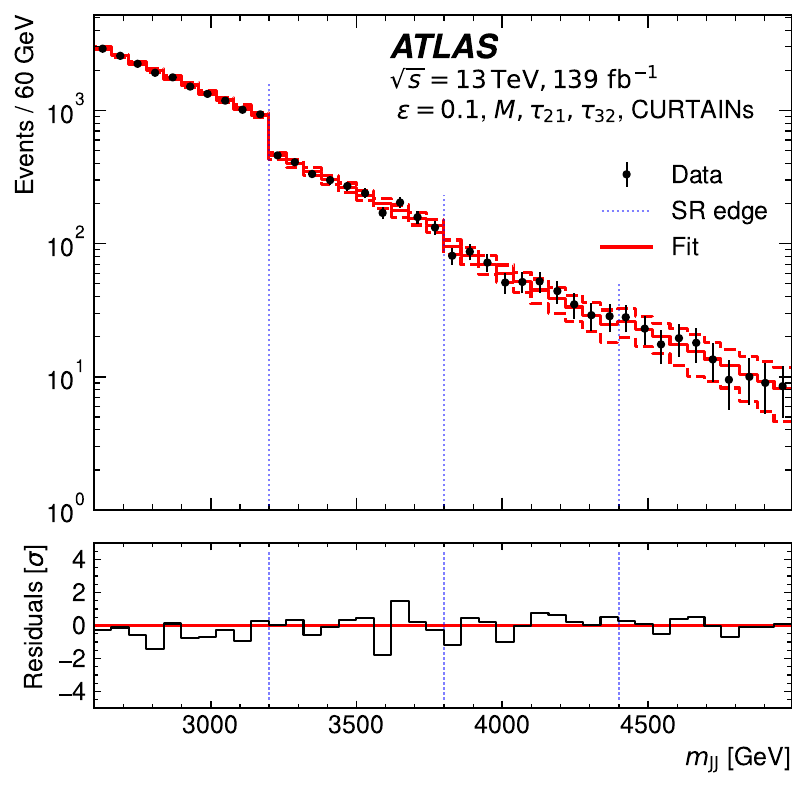}
\caption{}
\end{subfigure}
\caption{
Histograms of \mJJ in the first set of non-overlapping \mJJ SRs for the \Curtains method on all feature sets at the \loosewp classifier selection using the $\dY < 1.2$ data.
The figures show the resulting distributions when the analysis is evaluated on the first, third, fifth, and seventh \mJJ SRs.
Shown are different feature sets: (a) is the result of $\mathcal{T}=\monly$, (b) is the result of $\mathcal{T}=\mtau$ and (c) is the result of $\mathcal{T}=\mtautau$.
The fit is derived from the background-only fit interpolated from the SBs.
The uncertainties in the observed counts include the Poisson statistical uncertainty of the bin counts.
The uncertainties in the fit are represented by the dashed histograms and include the uncertainties in the fit parameters and the uncertainty from the classifier ensemble on the data.
The vertical dashed lines mark the edges of each SR in \mJJ.
The lower panel in each plot shows the Gaussian-equivalent significance of the deviation between the fit and data.
}
\label{fig:sig_hists_unblinded_body}
\end{figure}

This section presents the main results of the analysis.
\Cref{fig:sig_hists_unblinded_body} shows the distribution of events over \mJJ for different non-overlapping \mJJ SRs after the fit when using the \Curtains method.
Between neighboring SRs, sharp discontinuities are observed in the \mJJ spectrum.
These can appear because the classifier selection is performed in each SR independently.
Therefore, there is no guarantee that adjacent SRs will smoothly match each other.
A similar feature was observed in the previous ATLAS weakly supervised search~\cite{HDBS-2018-59}.

\begin{figure}[tb]
\centering
\begin{subfigure}[b]{0.45\textwidth}
\centering
\includegraphics[width=\textwidth]{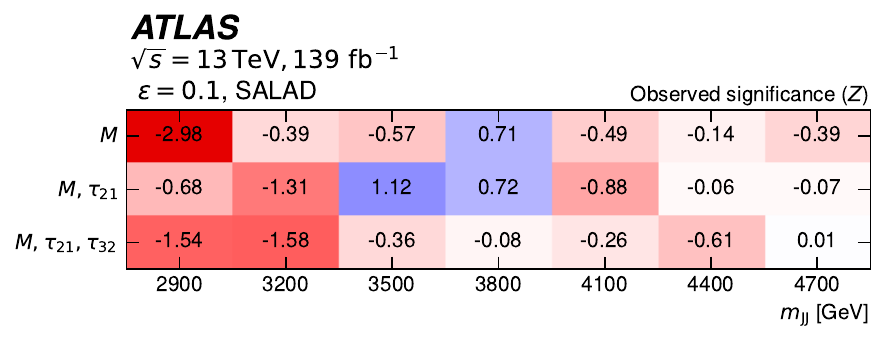}
\caption{}
\end{subfigure}
\hspace{1em} %
\begin{subfigure}[b]{0.45\textwidth}
\centering
\includegraphics[width=\textwidth]{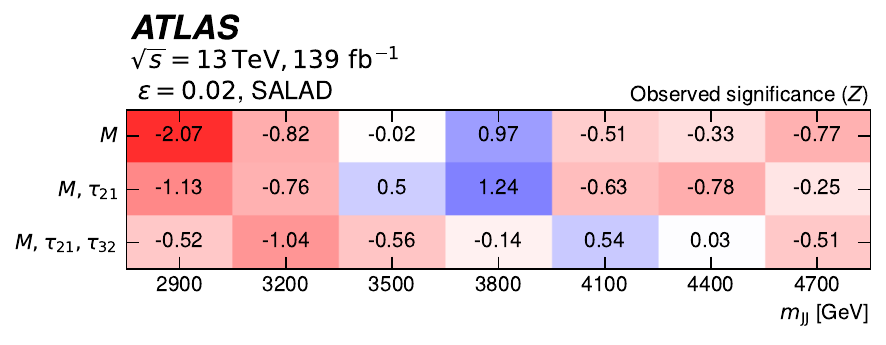}
\caption{}
\end{subfigure}
\caption{Observed significances ($Z$) for \Salad at the two different selections, (a) \loosewp and (b) \tightwp.
The significances are shown for all feature sets and \mJJ SRs.}
\label{fig:sig_tabl_unblinded salad}
\end{figure}

\begin{figure}[tb]
\centering
\begin{subfigure}[b]{0.45\textwidth}
\centering
\includegraphics[width=\textwidth]{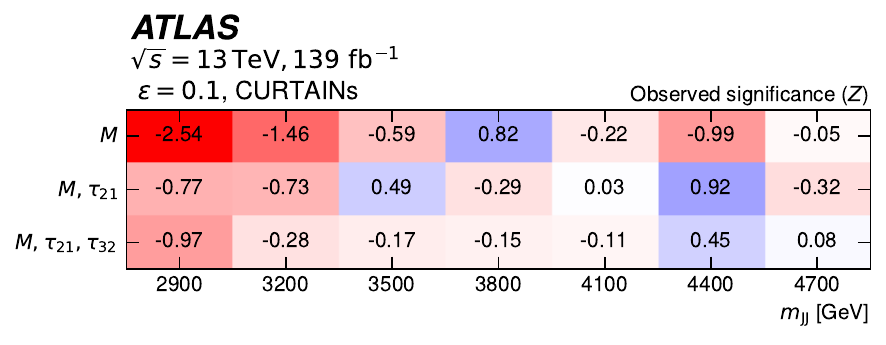}
\caption{}
\end{subfigure}
\hspace{1em} %
\begin{subfigure}[b]{0.45\textwidth}
\centering
\includegraphics[width=\textwidth]{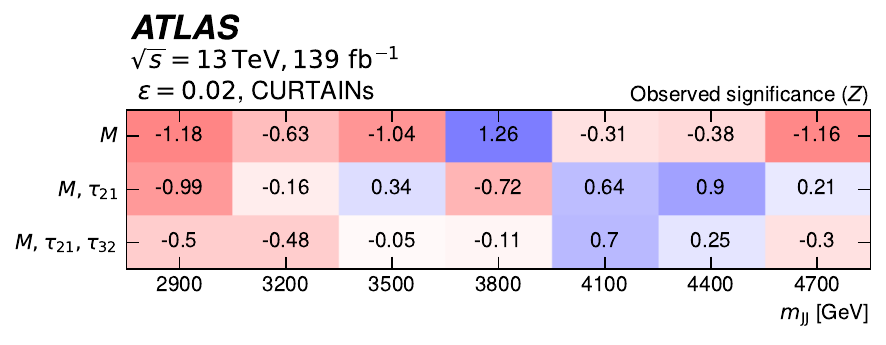}
\caption{}
\end{subfigure}
\caption{Observed significances ($Z$) for \Curtains at the two different selections, (a) \loosewp and (b) \tightwp.
The significances are shown for all feature sets and \mJJ SRs.}
\label{fig:sig_tabl_unblinded curtains}
\end{figure}

\Cref{fig:sig_tabl_unblinded salad} and \cref{fig:sig_tabl_unblinded curtains} show the significances of the observed data in the SRs for the \Salad and \Curtains methods, respectively.
The significances are extracted from the likelihood fit described in \cref{sec:likelihood fit}.
The largest observed excess for the \Salad (\Curtains) method has a local significance of $1.24 \sigma$ ($1.26 \sigma$).
There is a deficit in the first \mJJ SR with a local significance of $-2.98 \sigma$ ($-2.54 \sigma$) when using the \Salad (\Curtains) method and an efficiency of the \Cwola classifier of $\epsilon=0.1$.
This deficit was not observed on the test data sample as shown in \cref{app:testdatasetsvalidation:validation}.
For all other selections and \mJJ SRs both the  \Salad and \Curtains behave similarly to the results on the test data samples.

\subsection{Signal enhancement}
This section shows the ability of the analysis to enhance the potential signals, described in \cref{sec:data}, when they are injected into the SR\@.
For this test, into each of the regions 2.6 -- 3.2~\TeV and 4.1 -- 4.7~\TeV, a count of $S=3\sqrt{B}$ simulated signal events were injected, with $B$ the number of background events in each region.
Thus, the expected significance given perfect knowledge of the data distribution over \mJJ in the SR with no systematic uncertainty, would be $3\sigma$.
As the analysis incorporates information from more features than just \mJJ, this significance can be enhanced for some signals.

\begin{figure}[tb]
\centering
\begin{subfigure}[b]{\textwidth}
\centering
\includegraphics[width=0.75\textwidth]{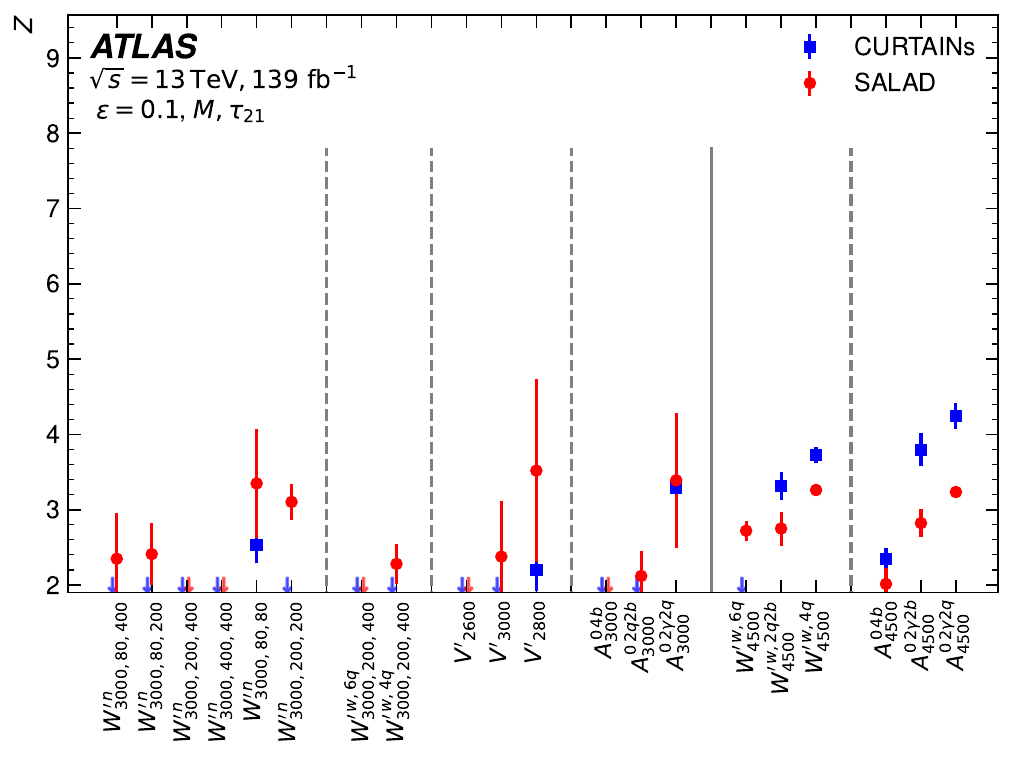}
\caption{}
\end{subfigure}
\begin{subfigure}[b]{\textwidth}
\centering
\includegraphics[width=0.75\textwidth]{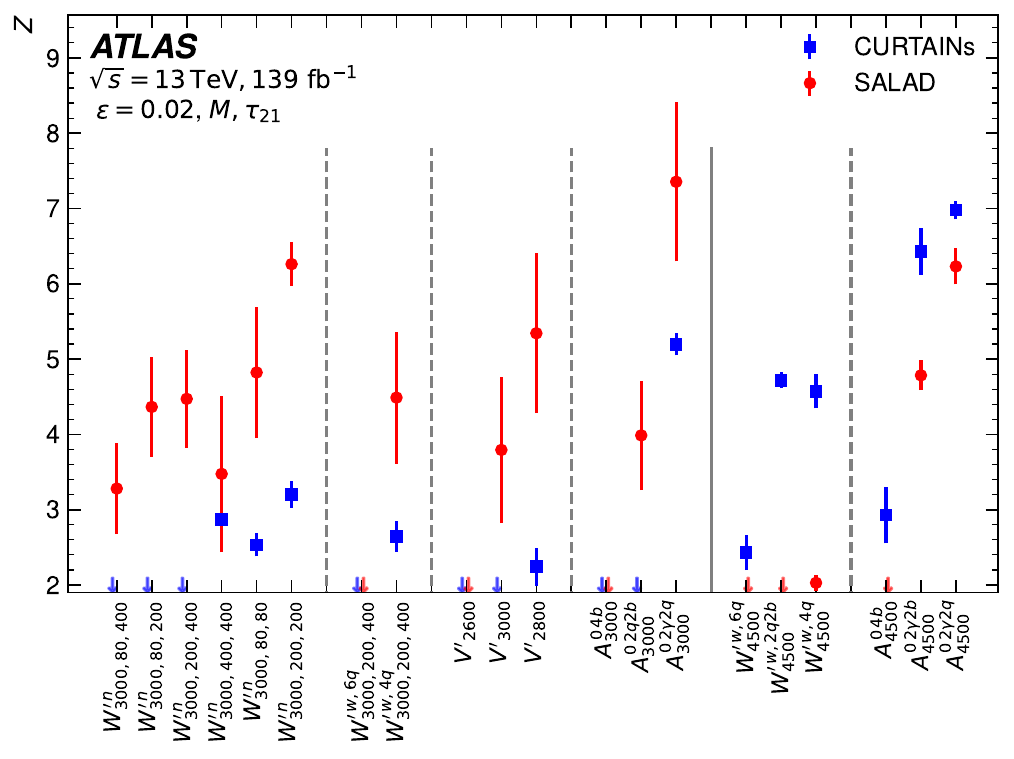}
\caption{}
\end{subfigure}
\caption{
Signal injection tests with $\mathcal{T}=\mtau$ at \loosewp in (a) and \tightwp in (b) for both the  \Salad and \Curtains for all simulated signal models.
The signal models are described in detail in \cref{sec:data}.
The local observed significance ($Z$) is shown as reported by the analysis pipeline after running the analysis with $3\sigma$ of signal injected into the data in the \mJJ SR centered on the different signals.
The errors show one standard deviation on the reported significance as calculated by bootstrapping the signal injection procedure.
}
\label{fig:signal_compare}
\end{figure}

In \Cref{fig:signal_compare},  the lower end of the \mJJ spectrum \Salad is better able to enhance signals than \Curtains, whereas the opposite is true at the higher end of the \mJJ spectrum.
The two methods are therefore complementary.
The tighter selection is also observed to better enhance signal across both the  methods.
Both the methods can be seen to return an observed significance $Z>3$ for the $3\sigma$ signal injection for many of the injected signals.
This reflects the significance enhancement ability of both the  \Salad and \Curtains.
The analysis may not be sensitive to all possible signal models, and this is what is observed.
The \mtau feature set is seen to have the broadest sensitivity to new physics.

\subsection{Exclusion limits}
\label{sec:exclusion}
This section shows the ability of the analysis to exclude the potential signals, described in \cref{sec:data}, using the full $CL_S$ prescription.
Tight limits can be placed on a wide range of signal models, which indicates the method's broad sensitivity.

\begin{figure}[tb]
\centering
\includegraphics[width=0.9\textwidth]{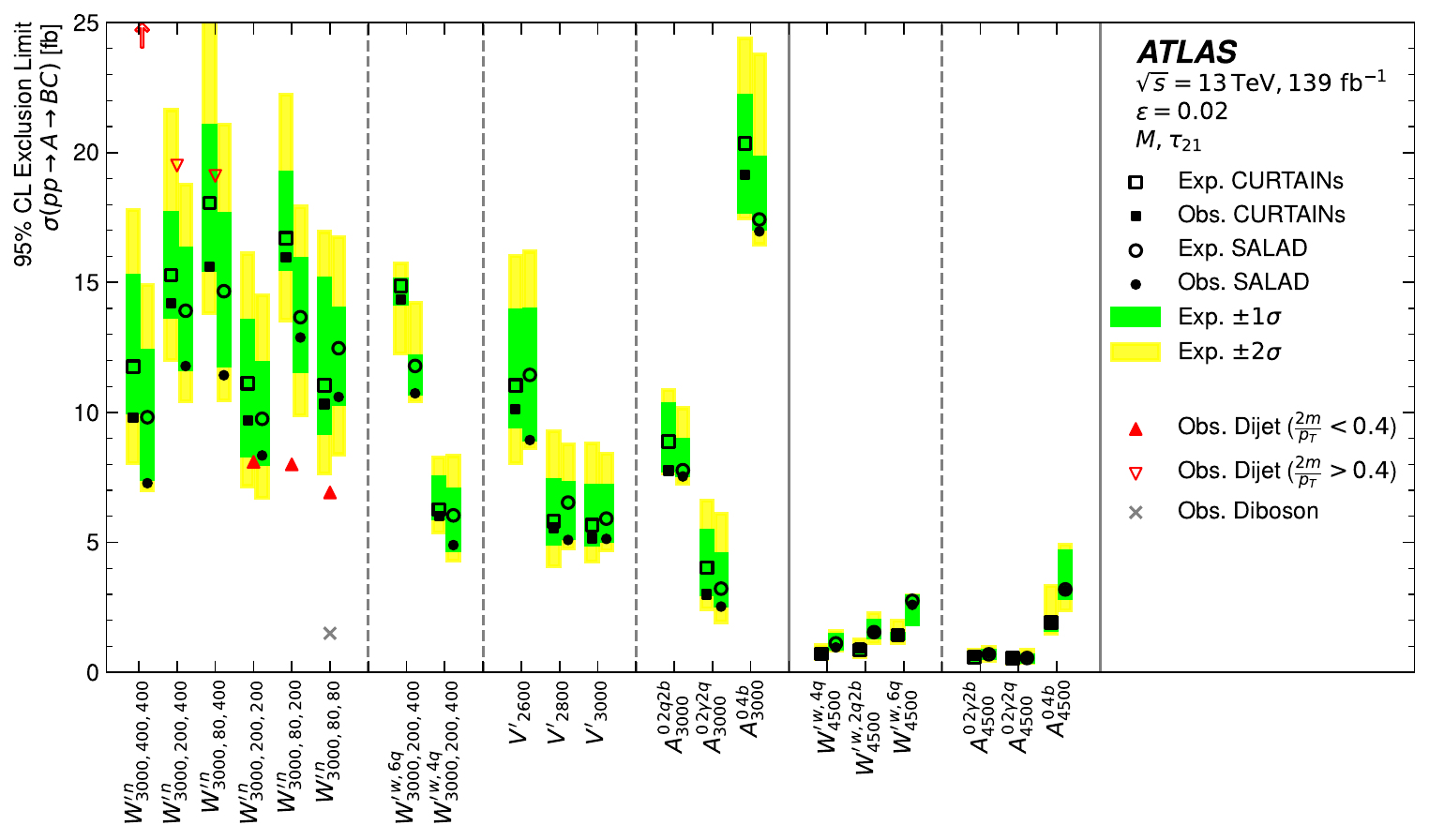}
\caption{Comparison of the 95\% CL upper limits on $\sigma(pp \rightarrow A \rightarrow BC)$ set by \Salad and \Curtains at \tightwp with $\mathcal{T}=\mtau$.
The one and two sigma variations on the expected limits for both the \Salad and \Curtains methods are shown as the shaded regions.
The signal models are  detailed in \cref{sec:data}.
The observed limits from the ATLAS dijet search~\cite{EXOT-2019-03} and the ATLAS all-hadronic diboson search~\cite{HDBS-2018-31} are shown in the red triangles and grey X-symbols respectively, and were derived in the previous weakly supervised ATLAS search~\cite{HDBS-2018-59}.
Limits for the inclusive dijet search are calculated using the $W'$ signals from this paper and the analysis of Ref.~\cite{EXOT-2019-03}; the diboson search limits are computed using the Heavy Vector Triplet~\cite{Pappadopulo_2014} $W'$ signal from Ref.~\cite{HDBS-2018-31}. The acceptance for the $W'$ in this paper, compared with the $W'$ acceptance in Ref.~\cite{HDBS-2018-31}, is $86\%$. The limit from the inclusive dijet search which exceeds the shown range is marked with a red arrow.}
\label{fig:curtains_salad_limits}
\end{figure}

\Cref{fig:curtains_salad_limits} compares the limits set by the analysis to previous dijet~\cite{EXOT-2019-03} and diboson~\cite{HDBS-2018-31} searches.
These classical limits are taken from Ref.~\cite{HDBS-2018-59} and were originally obtained by recasting the analysis with the new signal cases.
The new analysis sets stricter limits than the existing dijet search on signal models with daughter masses high enough such that $\frac{2m}{p_\mathrm{T}} > 0.4$ (signal models: $W'^{n}_{3000,400,400}$, $W'^{n}_{3000,200,400}$, $W'^{n}_{3000,80,400}$) and slightly weaker limits on signal cases with daughter masses such that $\frac{2m}{p_\mathrm{T}} < 0.4$ (signal models: $W'^{n}_{3000,200,200}$, $W'^{n}_{3000,80,200}$, $W'^{n}_{3000,80,80}$).
This is because the dijet search uses $R = 0.4$ jets, and therefore performs worse for cases where $\frac{2m}{p_\mathrm{T}} > 0.4$ and the resulting jets are too large to be contained inside the $R = 0.4$ jet radius.
The diboson search succeeds for signals with daughter masses close to the SM W mass $W'^{n}_{3000,80,80}$ but fails for other masses.

\begin{figure}[tb]
\centering
\includegraphics[width=0.9\textwidth]{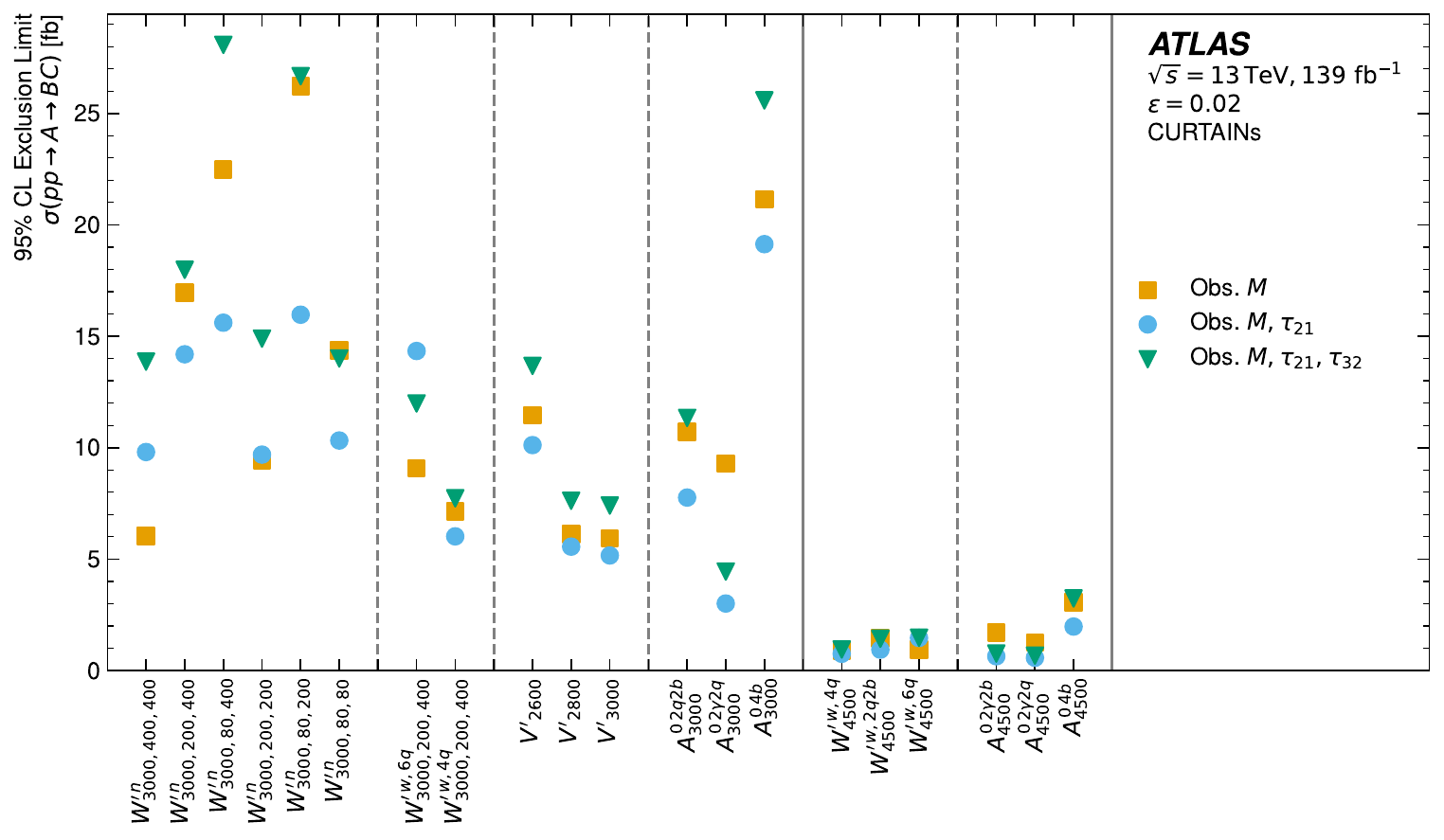}
\caption{Comparison of the 95\% CL upper limits on $\sigma(pp \rightarrow A \rightarrow BC)$ for different feature sets.
All limits use the \Curtains method with the \tightwp selection.
The signal models are  detailed in \cref{sec:data}.
}
\label{fig:curtains_feature_compare}
\end{figure}

In \Cref{fig:curtains_salad_limits}  both the  \Salad and \Curtains set similar limits on the signal cross-section at both the  ends of the \mJJ spectrum.
The same is true for all selections, feature sets and signals.
In \Cref{fig:curtains_feature_compare}  the \mtau feature set is able to set the strictest limits on the signal cross-section across almost all signals and SRs.
This is true for both the \Salad and \Curtains methods.
At the \loosewp selection in the $2.6-3.2~\TeV$ SR, the \monly feature set results in the strictest limits on the signal cross-section across almost all signals due to the deficit observed in \Cref{fig:sig_hists_unblinded_body}, which only appears in this region for this feature set.
Otherwise, the \mtau feature set most often sets the strictest limits in all regions and selections for both the  methods.

It is interesting  that the \mtautau feature set is not significantly better than the other two feature sets for topologies, like the BSM $W''/Z''\rightarrow qqq$ decays, that are expected to have a boosted 3-prong structure.
This may indicate that $\tau_{32}$ does not add enough information for these topologies to be useful for classification, and would be an interesting area for future study.

The analysis uses the full $CL_S$ method to set limits, and so direct comparisons cannot be made to the limits set by the previous weakly supervised dijet search which used a different approach~\cite{HDBS-2018-59}.
However, the previous weakly supervised search used the \monly feature set and so this should have similar sensitivity to the previous search.
Given that the \mtau feature set generally improves the sensitivity of the analysis, it represents an improvement on the previous weakly supervised search.
In a weakly supervised anomaly detection search, it is generally not known beforehand which features are most sensitive to a potential signal.
Additionally, as mentioned in \cref{sec:analysis}, including uninformative features can degrade the performance of the classifier.
The scan over many different feature sets is therefore one of the strengths of the analysis.

\FloatBarrier

\section{Conclusions}
\label{sec:conclusion}
A signal agnostic anomaly detection search is described for narrow-width resonances beyond the SM that produce a pair of jets.
The search uses 139 fb$^{-1}$ of $pp$ collisions at $\sqrt{s}=13$~\TeV recorded during 2015—2018 with the ATLAS detector at the LHC\@.
The analysis is designed to be sensitive to a broad range of new physics effects and targets resonances on the \mJJ spectrum between 2.6~\TeV to 5.0~\TeV.
After a selection through a weakly supervised classier, a bump hunt is performed to extract potential local excesses.
Two different strategies based on machine learning are used to create background models which are utilized in the classifier training.

The search extends the previous weakly supervised search performed by the ATLAS experiment in various ways.
The direct background model, generated using either the \Salad reweighting technique or the \Curtains morphing method, allows more features of the events to be used in the search than before.
Specifically, this search uses jet substructure features, which are related to the number of prongs in the jets, thereby particularly enhancing the sensitivity to boosted massive particles.
Additional benchmark signal models are integrated to probe the sensitivity of the search.
New validation methods are developed to address the challenge of developing a blind weakly supervised AD search.

Several signal models (the $W'^{n}$ family) were tested in both this analysis and the previous weakly supervised search by ATLAS. A direct comparison is not possible because of differences in the statistical treatment. Nevertheless, for some signals, the limits of the current analysis are of the same order of magnitude as in the previous search, if weaker by a factor of 2--3; while for the two signal models with the largest mass difference between the daughters ($W'^{n}_{3000,80,400}$ and $W'^{n}_{3000,80,200}$) the limits are improved by almost an order of magnitude.
In the present analysis, the \mtau feature set is shown to be broadly more sensitive than the \monly feature set, which corresponds to the training observables used in the previous analysis. This indicates that, should the previous search be repeated with the full $CL_S$ prescription, the limits would be closer to those observed in the current analysis.

No significant excess is observed in the data, consistent with the validation on the test data samples.
To illustrate the sensitivity of the analysis, upper limits at $95\%$ CL are set on the production cross-section for a set of signal benchmark models.
These limits are compared between the two background estimation methods and two existing searches.
The search is sensitive to many potential signal models which promote resonant dijet final states, such as BSM models including vector bosons $W'$ or pseudoscalar bosons $A^0$.
Unlike most searches, when a signal is injected into the data, the event selection changes, and so the entire process must be rerun every time a new signal is injected at a new cross-section.
The two background estimation methods, \Salad and \Curtains, provide comparable yet somewhat complementary sensitivity across the range of signal models considered.

There are various ways that the analysis could be improved in the future.
Higher-dimensional feature spaces may broaden the sensitivity.
This may require new strategies for training the classifiers that scale well with the number of features.
Additional approaches to generate the background model, including the combination of approaches, may also improve the sensitivity.
A detailed study of differences between data and simulation may also lead to new insight to address challenges that arise in data that are absent from previous phenomenological studies.
While the analysis is a significant extension of the previous search, hadronic final states at the LHC present a vast phase-space that may yet reveal signs of new physics with new AD searches in the future.

\section*{Acknowledgements}
%

% The next lines are included from the .//acknowledgements/Acknowledgements.tex input file
%
%

%
%

We thank CERN for the very successful operation of the LHC and its injectors, as well as the support staff at
CERN and at our institutions worldwide without whom ATLAS could not be operated efficiently.

The crucial computing support from all WLCG partners is acknowledged gratefully, in particular from CERN, the ATLAS Tier-1 facilities at TRIUMF/SFU (Canada), NDGF (Denmark, Norway, Sweden), CC-IN2P3 (France), KIT/GridKA (Germany), INFN-CNAF (Italy), NL-T1 (Netherlands), PIC (Spain), RAL (UK) and BNL (USA), the Tier-2 facilities worldwide and large non-WLCG resource providers. Major contributors of computing resources are listed in Ref.~\cite{ATL-SOFT-PUB-2025-001}.

We gratefully acknowledge the support of ANPCyT, Argentina; YerPhI, Armenia; ARC, Australia; BMWFW and FWF, Austria; ANAS, Azerbaijan; CNPq and FAPESP, Brazil; NSERC, NRC and CFI, Canada; CERN; ANID, Chile; CAS, MOST and NSFC, China; Minciencias, Colombia; MEYS CR, Czech Republic; DNRF and DNSRC, Denmark; IN2P3-CNRS and CEA-DRF/IRFU, France; SRNSFG, Georgia; BMFTR, HGF and MPG, Germany; GSRI, Greece; RGC and Hong Kong SAR, China; ICHEP and Academy of Sciences and Humanities, Israel; INFN, Italy; MEXT and JSPS, Japan; CNRST, Morocco; NWO, Netherlands; RCN, Norway; MNiSW, Poland; FCT, Portugal; MNE/IFA, Romania; MSTDI, Serbia; MSSR, Slovakia; ARIS and MVZI, Slovenia; DSI/NRF, South Africa; MICIU/AEI, Spain; SRC and Wallenberg Foundation, Sweden; SERI, SNSF and Cantons of Bern and Geneva, Switzerland; NSTC, Taipei; TENMAK, T\"urkiye; STFC/UKRI, United Kingdom; DOE and NSF, United States of America.

Individual groups and members have received support from BCKDF, CANARIE, CRC and DRAC, Canada; CERN-CZ, FORTE and PRIMUS, Czech Republic; COST, ERC, ERDF, Horizon 2020, ICSC-NextGenerationEU and Marie Sk{\l}odowska-Curie Actions, European Union; Investissements d'Avenir Labex, Investissements d'Avenir Idex and ANR, France; DFG and AvH Foundation, Germany; Herakleitos, Thales and Aristeia programmes co-financed by EU-ESF and the Greek NSRF, Greece; BSF-NSF and MINERVA, Israel; NCN and NAWA, Poland; La Caixa Banking Foundation, CERCA Programme Generalitat de Catalunya and PROMETEO and GenT Programmes Generalitat Valenciana, Spain; G\"{o}ran Gustafssons Stiftelse, Sweden; The Royal Society and Leverhulme Trust, United Kingdom.

In addition, individual members wish to acknowledge support from Armenia: Yerevan Physics Institute (FAPERJ); CERN: European Organization for Nuclear Research (CERN DOCT); Chile: Agencia Nacional de Investigaci\'on y Desarrollo (FONDECYT 1230812, FONDECYT 1240864, Fondecyt 3240661); China: Chinese Ministry of Science and Technology (MOST-2023YFA1605700, MOST-2023YFA1609300), National Natural Science Foundation of China (NSFC - 12175119, NSFC 12275265, NSFC-12075060); Czech Republic: Czech Science Foundation (GACR - 24-11373S), Ministry of Education Youth and Sports (ERC-CZ-LL2327, FORTE CZ.02.01.01/00/22\_008/0004632), PRIMUS Research Programme (PRIMUS/21/SCI/017); EU: H2020 European Research Council (ERC - 101002463); European Union: European Research Council (BARD No. 101116429, ERC - 948254, ERC 101089007), Horizon 2020 Framework Programme (MUCCA - CHIST-ERA-19-XAI-00), European Union, Future Artificial Intelligence Research (FAIR-NextGenerationEU PE00000013), Italian Center for High Performance Computing, Big Data and Quantum Computing (ICSC, NextGenerationEU); France: Agence Nationale de la Recherche (ANR-20-CE31-0013, ANR-21-CE31-0013, ANR-21-CE31-0022, ANR-22-EDIR-0002); Germany: Baden-Württemberg Stiftung (BW Stiftung-Postdoc Eliteprogramme), Deutsche Forschungsgemeinschaft (DFG - 469666862, DFG - CR 312/5-2); Italy: Istituto Nazionale di Fisica Nucleare (ICSC, NextGenerationEU), Ministero dell'Università e della Ricerca (NextGenEU 153D23001490006 M4C2.1.1, NextGenEU I53D23000820006 M4C2.1.1, NextGenEU I53D23001490006 M4C2.1.1); Japan: Japan Society for the Promotion of Science (JSPS KAKENHI JP22H01227, JSPS KAKENHI JP22H04944, JSPS KAKENHI JP22KK0227, JSPS KAKENHI JP23KK0245); Norway: Research Council of Norway (RCN-314472); Poland: Ministry of Science and Higher Education (IDUB AGH, POB8, D4 no 9722), Polish National Agency for Academic Exchange (PPN/PPO/2020/1/00002/U/00001), Polish National Science Centre (NCN 2021/42/E/ST2/00350, NCN OPUS 2023/51/B/ST2/02507, NCN OPUS nr 2022/47/B/ST2/03059, NCN UMO-2019/34/E/ST2/00393, UMO-2020/37/B/ST2/01043, UMO-2021/40/C/ST2/00187, UMO-2022/47/O/ST2/00148, UMO-2023/49/B/ST2/04085, UMO-2023/51/B/ST2/00920); Portugal: Foundation for Science and Technology (FCT); Spain: Generalitat Valenciana (Artemisa, FEDER, IDIFEDER/2018/048), Ministry of Science and Innovation (MCIN \& NextGenEU PCI2022-135018-2, MICIN \& FEDER PID2021-125273NB, RYC2019-028510-I, RYC2020-030254-I, RYC2021-031273-I, RYC2022-038164-I); Sweden: Carl Trygger Foundation (Carl Trygger Foundation CTS 22:2312), Swedish Research Council (Swedish Research Council 2023-04654, VR 2018-00482, VR 2021-03651, VR 2022-03845, VR 2022-04683, VR 2023-03403), Knut and Alice Wallenberg Foundation (KAW 2018.0458, KAW 2019.0447, KAW 2022.0358); Switzerland: Swiss National Science Foundation (SNSF - PCEFP2\_194658); United Kingdom: Leverhulme Trust (Leverhulme Trust RPG-2020-004), Royal Society (NIF-R1-231091); United States of America: U.S. Department of Energy (ECA DE-AC02-76SF00515), Neubauer Family Foundation.

%
%

% End of text imported from the .//acknowledgements/Acknowledgements.tex input file

%
%
%
%
%
%
%

%
\clearpage
\appendix
\part*{Appendix}
\addcontentsline{toc}{part}{Appendix}
%

%
%

% The next lines are included from the .//appendices/implementation/main.tex input file
\section{Analysis implementation}
\label{sec:implementation_appendix}
All classifiers are sequential neural networks with four layers of size 64, 64, 64, 1 and activation functions \texttt{ReLu}, \texttt{ReLu}, \texttt{ReLu}, sigmoid, respectively.
A dropout of 5\% was used between each layer to reduce overfitting.
These models were implemented in \texttt{Keras}~\cite{chollet2015keras} using the \texttt{Tensorflow}~\cite{Abadi:2016kxi} backend.
The classifiers use binary cross-entropy loss, and training loss is minimized using the \texttt{Adam}~\cite{kingma2014adam} optimizer with a learning rate of 0.001 and no scheduling.
For the reweighting classifier, half of the data was used for training, and the other half for validation.
Training proceeded for 200 epochs with early stopping (with 25-epoch patience) on the validation loss and a batch size of 512.

All \Curtains models used the same settings as those used in Ref.~\cite{Raine:2022hht}.
All normalizing flows were implemented using the \texttt{nflows} package~\cite{nflows} in \texttt{pytorch}~\cite{Pytorch}.
All flows use rational-quadratic spline layers~\cite{durkan2019neural}.

For the signal-tagging classifier, the data was split into five tranches.
five classifiers were then trained in a five-fold validation scheme --- each classifier was trained on 3 tranches of the data, validated on one tranche, and evaluated on the remaining tranche.
The classifiers used the same architecture as the reweighting classifier, but with a softmax activation function and binary cross-entropy loss.
Training proceeded for 100 epochs with a learning rate of 0.001, no learning rate scheduling, the \texttt{Adam} optimizer, batch sizes of 512, and early stopping with a patience of 10.

No hyperparameter optimization was performed for any of the models used  --- this would be an interesting area for future study. The optimization problem is nontrivial in the weakly supervised case, as a decision must be made about what to optimize.
The optimization is also difficult to do in a way that is signal model agnostic, as the signal model is not known a priori.

All methods used \texttt{numpy}~\cite{harris2020array}, \texttt{pandas}~\cite{the_pandas_development_team_2023_7857418}, \texttt{scipy}~\cite{scipy}, \texttt{matplotlib}~\cite{matplotlib}, \texttt{scikit-learn}~\cite{scikit-learn}, \texttt{scikit-hep}~\cite{Rodrigues:2020syo}, and pipeline runs were configured using \texttt{hydra}~\cite{Yadan2019Hydra} and \texttt{omegaconf}~\cite{omegaconf}.
The analysis was implemented as a workflow using \texttt{snakemake}~\cite{snakemake}.

% End of text imported from the .//appendices/implementation/main.tex input file

% The next lines are included from the .//appendices/validation/main.tex input file
\section{Test data samples}
\label{app:testdatasetsvalidation:datasets}
This section describes the three complementary and orthogonal test data samples that are used in this analysis.
Their construction is detailed in this section.
Each test data sample models the events in the SR of the analysis in a different way.
The \downupdataset test data sample is used in the non-closure correction described in \cref{sec:nonclosure}.
All three are used for the validation of the analysis described in \cref{sec:validation}.

\subsection{\deltaydataset test data sample}
In the SR, the absolute rapidity difference $\dY= \abs{y_1-y_2}$ between the two jets is required to satisfy $\dY < 1.2$.
A test data sample can be constructed by inverting this selection requirement to $\dY > 1.2$, referred to as \deltaydataset.
This data sample is expected to be dominated by $t$-channel QCD background events.
Several observables correlate with \dY, so the correlations in this data sample do not perfectly mimic those of the SR.

\subsection{\mcdataset test data sample}
A QCD dijet MC simulated sample can be used as a signal-free test data sample.
Exactly the same selection requirements can be applied to the MC simulated sample as are applied in data.
However, mismodeling and undersampling in the MC simulation can limit the validity of this test data sample.

To increase the sample size of the \mcdataset test data sample, a generative model is used.
First, an \mJJ conditional generative model is trained on a subset of the data, second, the samples are simulated using the true \mJJ samples as detailed below.
This strategy relies on the assumption that any signal, if present, is rare -- otherwise it would have been observed by previous model-agnostic searches like the generic dijet search~\cite{EXOT-2019-03,CMS-EXO-19-012}.
The performance of the weakly supervised method on this synthetic test data sample is expected to be similar to its performance on the real data, as both data samples cover the same kinematic phase-space.

Conditional diffusion models are used to learn the distribution of $p(x|m_{\mathrm{JJ}})$.
These diffusion models use continuous flow matching~\cite{lipman2023flow} with logitnorm time sampling~\cite{esser2024scaling}.
The diffusion networks are sequential neural networks with 3 layers of size 32, 32, 32 and activation functions \texttt{ReLu}, \texttt{ReLu}, \texttt{ReLu}, respectively.
The output layer has no activation function.
The models are trained using the \texttt{Adam} optimizer~\cite{kingma2014adam} with a learning rate of $10^{-4}$ and a batch size of 64 for 100 epochs.
The learning rate is linearly ramped up to $10^{-4}$ over the first 500 steps.

A conditional diffusion model was trained on the full MC simulated sample in the $\dY < 1.2$ region.
Samples are drawn from this model using \mJJ samples from data in the $\dY < 1.2$ region.
The \mJJ samples are not expected to be sensitive to resonances; this was studied by a previous dijet search~\cite{EXOT-2019-03}.
When upsampling the background only distribution using \mJJ samples from the $\dY < 1.2$ data, the resulting distribution is expected to be representative of the background in the $\dY < 1.2$ region.
Multiple diffusion models with different random seeds were trained on the MC simulated sample to provide some variability in the different upsampled data samples.

\subsection{\downupdataset test data sample}
The fully inclusive $\dY < 1.2$ data is not expected to contain any resonant new physics produced at large cross sections; such signals would likely have been seen in prior searches.
Thus, any signal present is assumed to be generated with a small cross section.
Randomly selecting a small fraction of events will further dilute the significance of any present signal, rendering the sample effectively blinded.

\begin{figure}[tb]
\centering
\includegraphics[width=0.5\textwidth]{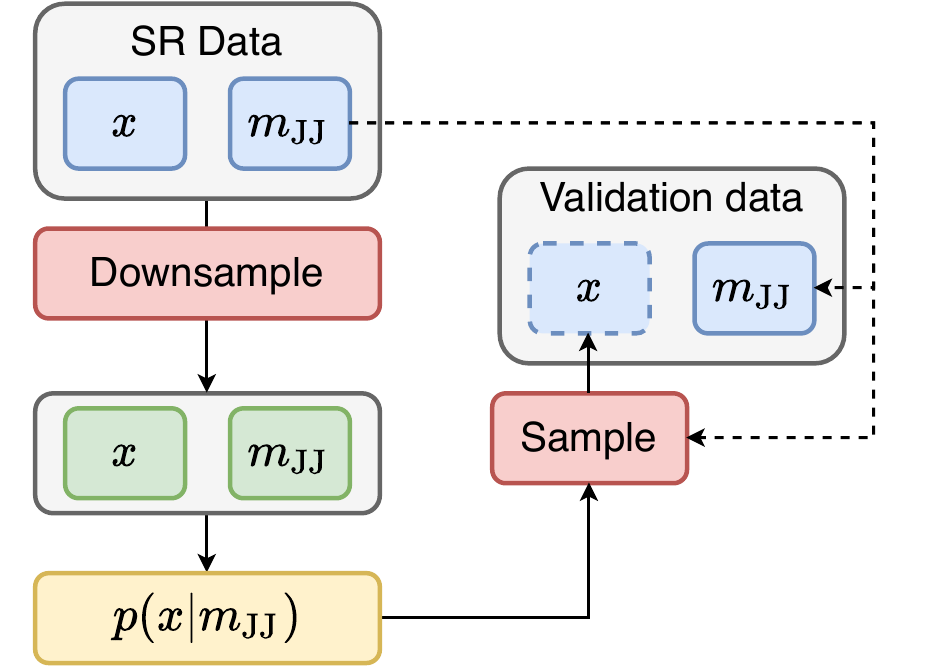}
\caption{Outline of the procedure used to construct the downsampled $\dY < 1.2$ data and upsampled validation sets.}
\label{fig:downsampled_validation_procedure}
\end{figure}

This randomly downsampled data sample does not have the correct statistics to be used as a test data sample, but it can be used to fit a conditional diffusion model $p(x|m_{\mathrm{JJ}})$ that can be upsampled using the fully inclusive \mJJ samples.
The upsampling follows the same strategy as outlined for the \mcdataset test data sample above.
The use of the \mJJ samples in the $\dY < 1.2$ region is justified similarly to the \mcdataset test data sample.
The whole procedure is graphically depicted in \cref{fig:downsampled_validation_procedure}.
The resulting diffusion model can generate samples that are representative of the data in the $\dY < 1.2$ region but lack sensitivity to new physics.
Conditional diffusion models are also smooth and do not introduce resonance-like artifacts in the conditional distribution.

This approach is validated in the $\dY > 1.2$ region by injecting a signal into the inclusive data sample such that the analysis produced an excess with a significance of $Z=3\sigma$ in the $2.6-3.2~\TeV$ \mJJ SR.
Randomly downsampling this data sample, then upsampling resulted in no excess appearing in any SR.
This confirms that the \downupdataset procedure eliminates sensitivity to injected signals and does not introduce artifacts that produce a resonance-like effect in \mJJ.

\section{Validation}
\label{app:testdatasetsvalidation:validation}
This section explains the validation strategy and shows the results of the validation of the analysis in greater detail.
It uses all three test data samples described in \cref{app:testdatasetsvalidation:datasets}.
The \downupdataset test data sample sets the strictest requirements for the validation.

\begin{figure}[tb]
\centering
\begin{subfigure}[b]{0.3\textwidth}
\centering
\includegraphics[width=\textwidth]{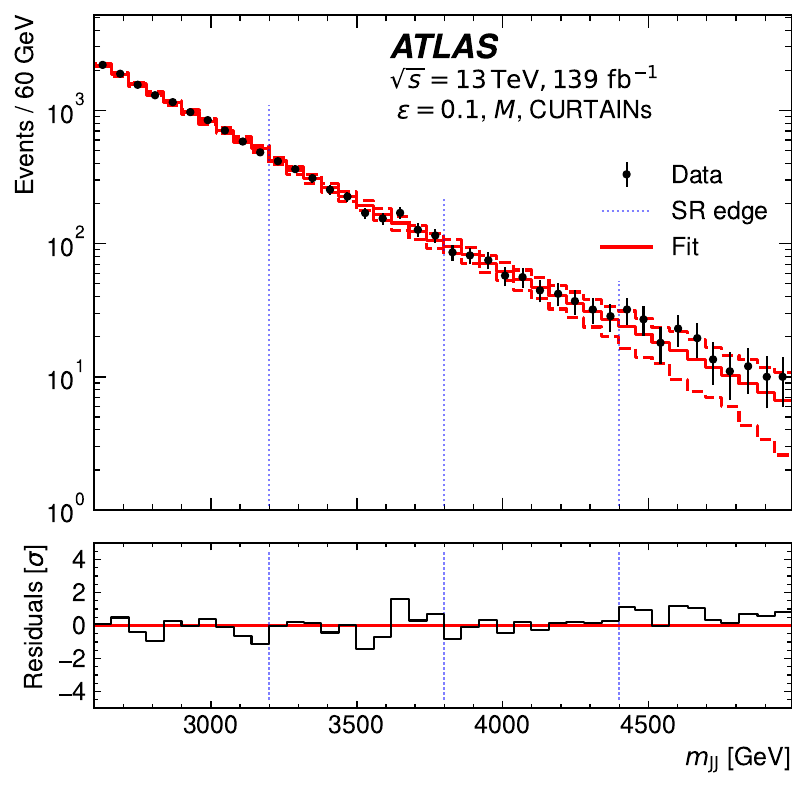}
\caption{}
\end{subfigure}
\hspace{0.01\textwidth} %
\centering
\begin{subfigure}[b]{0.3\textwidth}
\centering
\includegraphics[width=\textwidth]{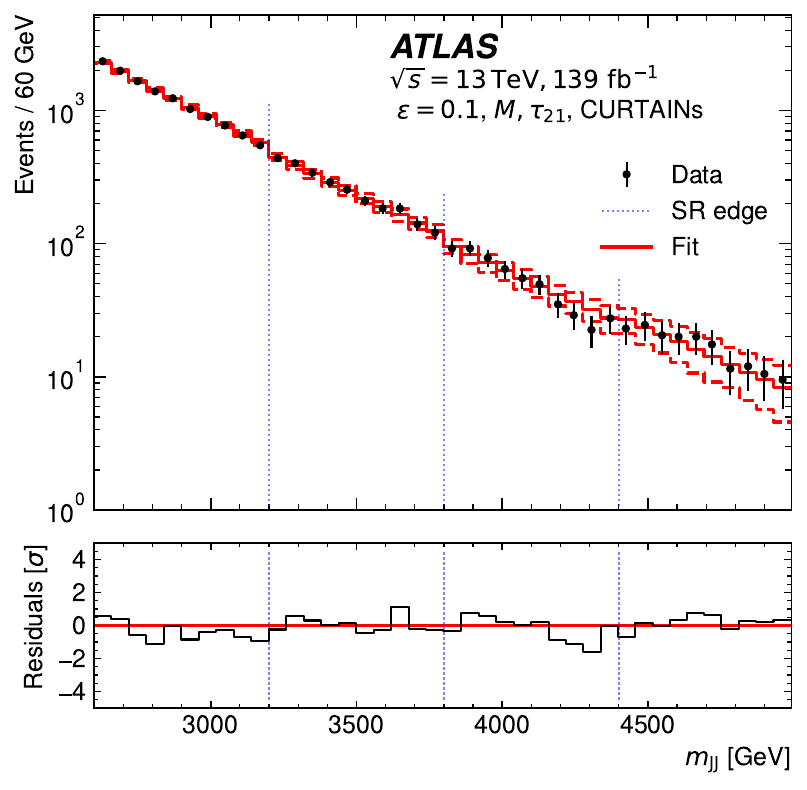}
\caption{}
\end{subfigure}
\hspace{0.01\textwidth} %
\centering
\begin{subfigure}[b]{0.3\textwidth}
\centering
\includegraphics[width=\textwidth]{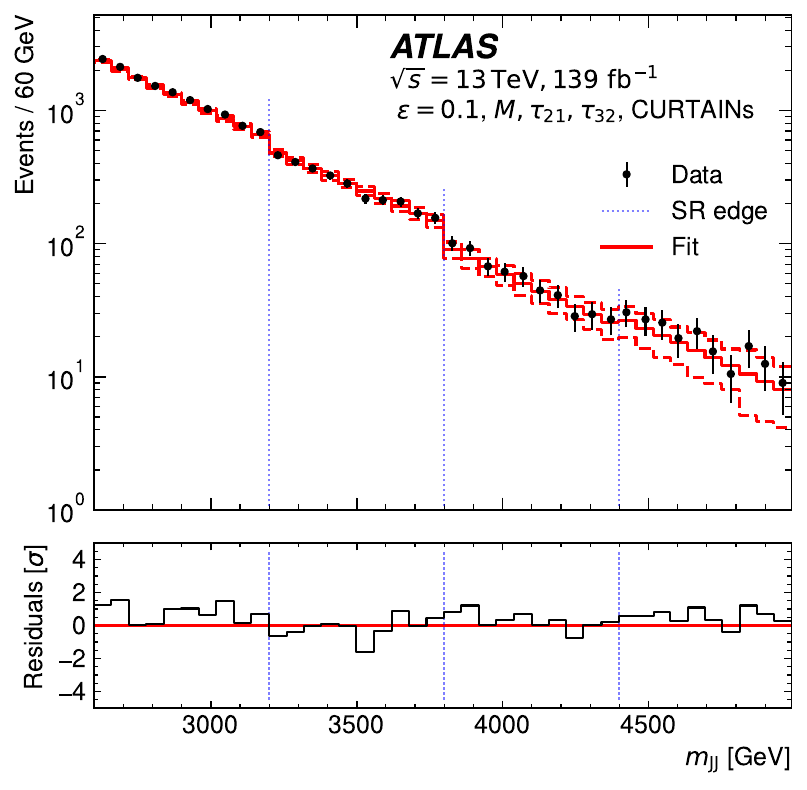}
\caption{}
\end{subfigure}
\caption{
Histograms of \mJJ in the first set of non-overlapping \mJJ SRs for the \Curtains method on all feature sets at the \loosewp classifier selection on one \downupdataset test data sample.
The figures show the resulting distributions when the analysis is evaluated on the first, third, fifth, and seventh \mJJ SRs.
Shown are different feature sets: (a) is the result of $\mathcal{T}=\monly$, (b) is the result of $\mathcal{T}=\mtau$ and (c) is the result of $\mathcal{T}=\mtautau$.
The fit is derived from the background-only fit interpolated from the SBs.
The uncertainties in the observed counts include the Poisson statistical uncertainty of the bin counts.
The uncertanties on the fit are represented by the dashed histograms and include the uncertainties in the fit parameters and the uncertainty from the classifier ensemble on the data.
The vertical dashed lines mark the edges of each SR in \mJJ.
The lower panel in each plot shows the Gaussian-equivalent significance of the deviation between the fit and data.
}
\label{fig:sig_hists_validation_appendix}
\end{figure}

\Cref{fig:sig_hists_validation_appendix} shows the distribution of events over \mJJ for different non-overlapping \mJJ SRs after the fit when using the \Curtains method, in one instance of the \downupdataset data sample.
It shows that the observations are consistent with the prediction of the background-only fit interpolated from the SBs.
The same is true in all SRs for both the methods and all test data samples.

\begin{figure}[tb]
\centering
\begin{subfigure}[b]{0.45\textwidth}
\centering
\includegraphics[width=\textwidth]{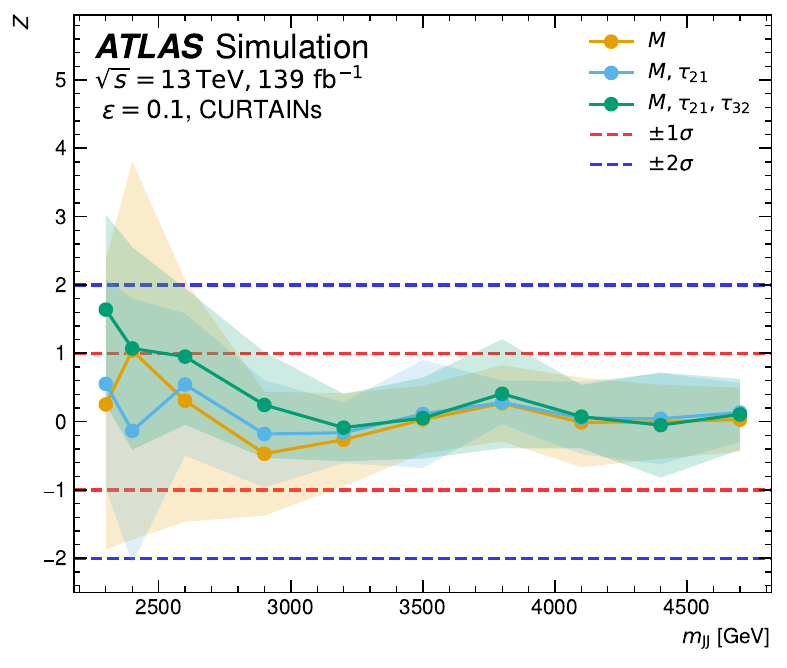}
\caption{}
\end{subfigure}
\begin{subfigure}[b]{0.45\textwidth}
\centering
\includegraphics[width=\textwidth]{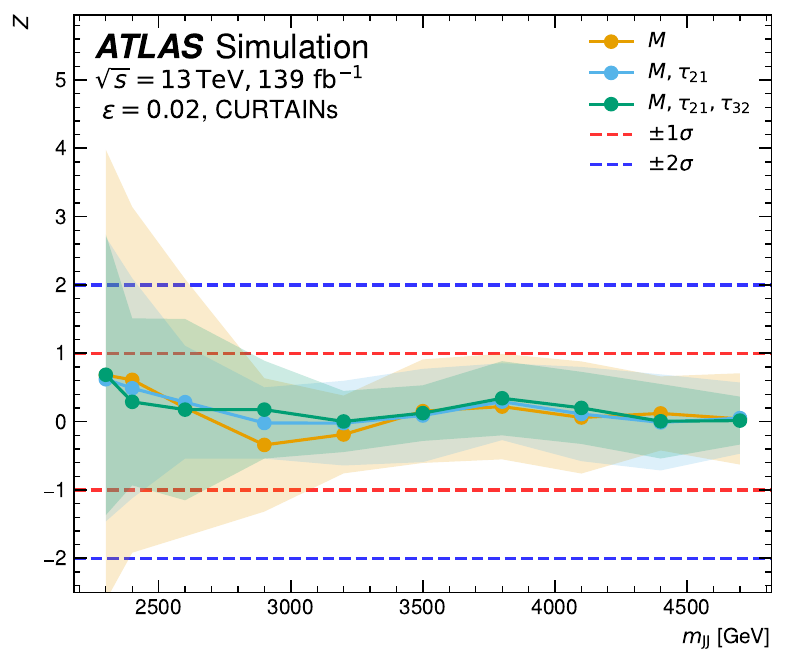}
\caption{}
\end{subfigure}
\\
\begin{subfigure}[b]{0.45\textwidth}
\centering
\includegraphics[width=\textwidth]{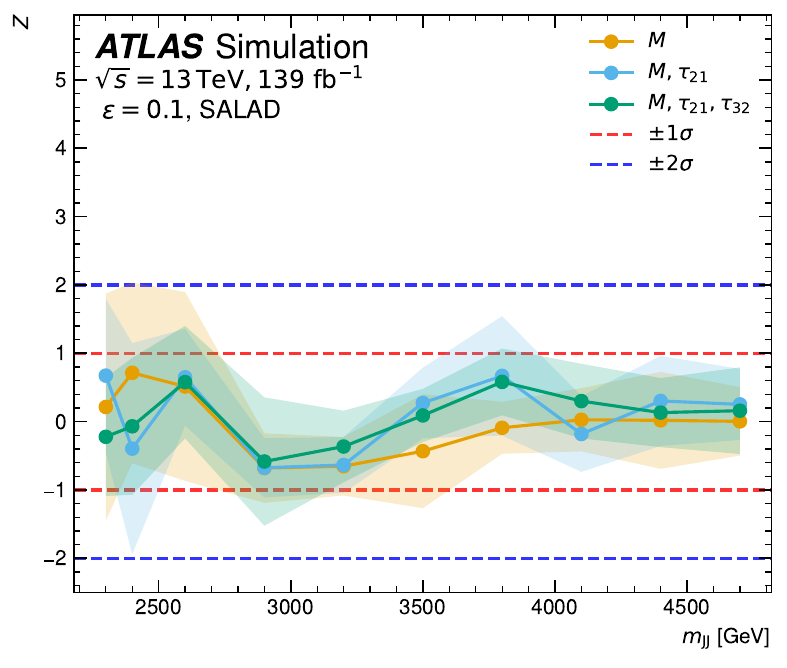}
\caption{}
\end{subfigure}
\begin{subfigure}[b]{0.45\textwidth}
\centering
\includegraphics[width=\textwidth]{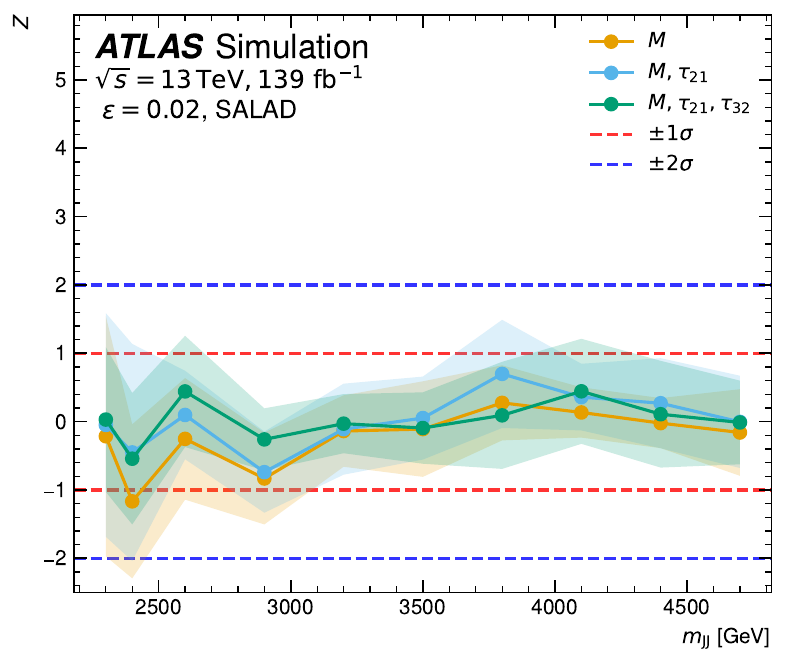}
\caption{}
\end{subfigure}
\caption{Central values and spread of the local significances (Z) for (a, b) \Curtains and (c, d) \Salad at the two different selections, (a, c) \loosewp and (b, d) \tightwp.
Significances are shown for all feature sets and \mJJ SR centers.
The bands include the distributions of ten different instances of the \mcdataset test data sample.}
\label{fig:val_sig_spread_neg_mc}
\end{figure}

\begin{figure}[tb]
\centering
\begin{subfigure}[b]{0.45\textwidth}
\centering
\includegraphics[width=\textwidth]{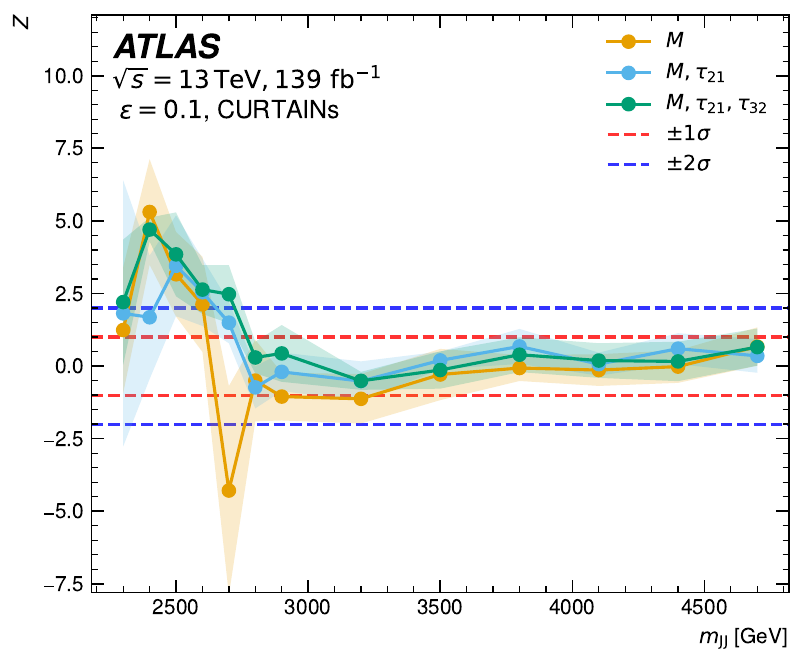}
\caption{}
\end{subfigure}
\begin{subfigure}[b]{0.45\textwidth}
\centering
\includegraphics[width=\textwidth]{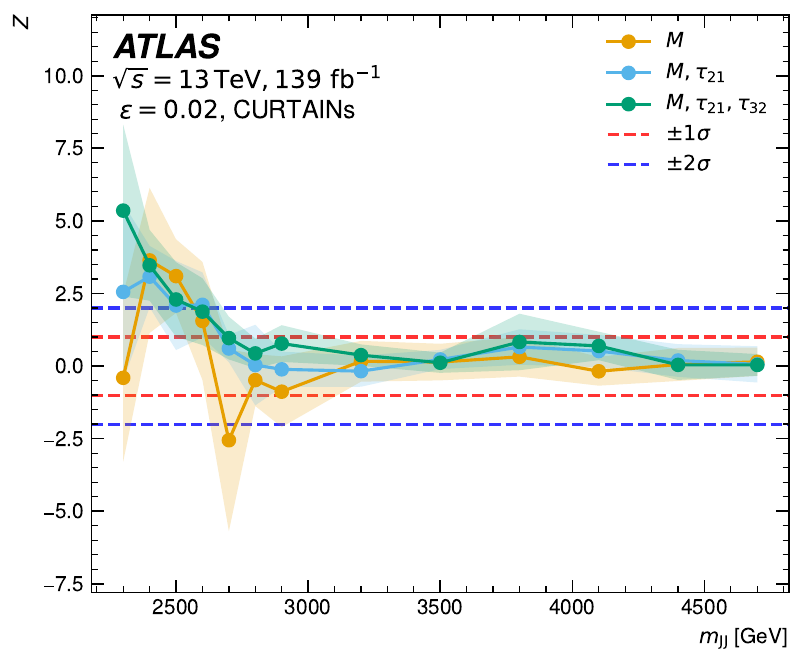}
\caption{}
\end{subfigure}
\\
\begin{subfigure}[b]{0.45\textwidth}
\centering
\includegraphics[width=\textwidth]{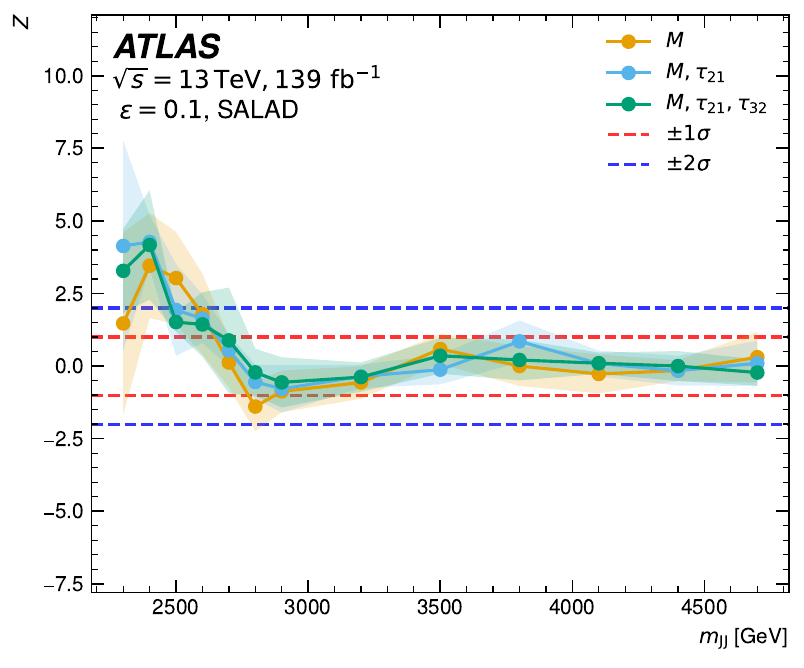}
\caption{}
\end{subfigure}
\begin{subfigure}[b]{0.45\textwidth}
\centering
\includegraphics[width=\textwidth]{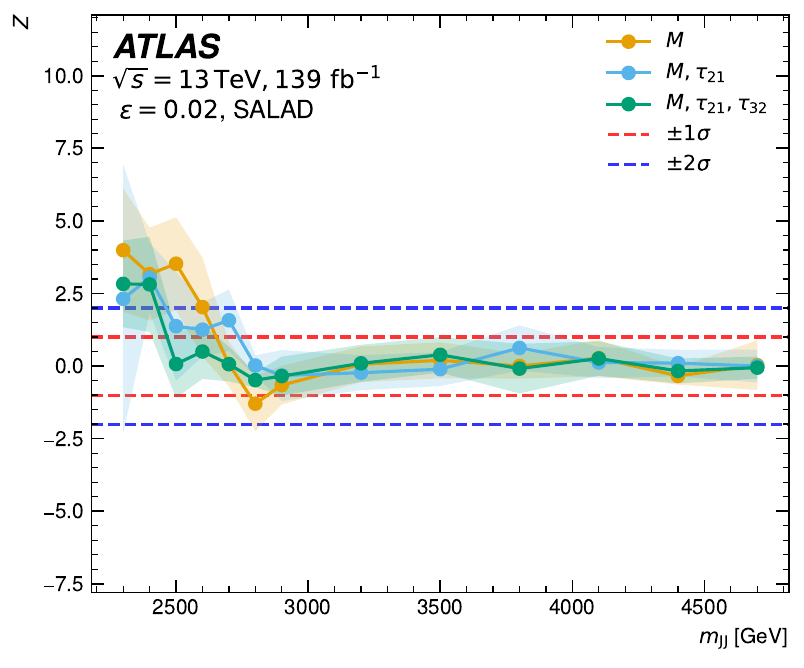}
\caption{}
\end{subfigure}
\caption{Central values and spread of the local significances (Z) for (a, b) \Curtains and (c, d) \Salad at the two different selections, (a, c) \loosewp and (b, d) \tightwp.
Significances are shown for all feature sets and \mJJ SR centers.
The bands include the distributions of ten different instances of the \downupdataset test data sample.}
\label{fig:val_sig_spread_neg_down_up}
\end{figure}

\Cref{fig:val_sig_spread_neg_mc} (\Cref{fig:val_sig_spread_neg_down_up}) shows the mean and spread of the significance of the data above the background estimate for \Curtains and \Salad at the two different selections, \loosewp and \tightwp as a function of \mJJ across ten different instances of the \mcdataset (\downupdataset) test data sample.
These distributions are used to select the \mJJ SRs that passed the validation and are used in the final analysis according to the description in \cref{sec:validation}.

% End of text imported from the .//appendices/validation/main.tex input file

\clearpage

\printbibliography
\clearpage
% ATLAS Collaboration author list
% Reference date of HMBS-2024-34 is 2024-11-06
% Author list last updated on date 08-SEP-25
% Data extracted on 08-Sep-2025 for paper reference HMBS-2024-34
% at 11:48am
 
\begin{flushleft}
\hypersetup{urlcolor=black}
{\Large The ATLAS Collaboration}

\bigskip

\AtlasOrcid[0000-0002-6665-4934]{G.~Aad}$^\textrm{\scriptsize 105}$,
\AtlasOrcid[0000-0001-7616-1554]{E.~Aakvaag}$^\textrm{\scriptsize 17}$,
\AtlasOrcid[0000-0002-5888-2734]{B.~Abbott}$^\textrm{\scriptsize 124}$,
\AtlasOrcid[0000-0002-0287-5869]{S.~Abdelhameed}$^\textrm{\scriptsize 120a}$,
\AtlasOrcid[0000-0002-1002-1652]{K.~Abeling}$^\textrm{\scriptsize 56}$,
\AtlasOrcid[0000-0001-5763-2760]{N.J.~Abicht}$^\textrm{\scriptsize 50}$,
\AtlasOrcid[0000-0002-8496-9294]{S.H.~Abidi}$^\textrm{\scriptsize 30}$,
\AtlasOrcid[0009-0003-6578-220X]{M.~Aboelela}$^\textrm{\scriptsize 46}$,
\AtlasOrcid[0000-0002-9987-2292]{A.~Aboulhorma}$^\textrm{\scriptsize 36e}$,
\AtlasOrcid[0000-0001-5329-6640]{H.~Abramowicz}$^\textrm{\scriptsize 158}$
% ...
\bigskip
\\

% ...
$^{17}$Department for Physics and Technology, University of Bergen, Bergen; Norway.\\
% ...
$^{30}$Physics Department, Brookhaven National Laboratory, Upton NY; United States of America.\\
% ...
$^{36}$$^{(a)}$Facult\'e des Sciences Ain Chock, Universit\'e Hassan II de Casablanca;$^{(b)}$Facult\'{e} des Sciences, Universit\'{e} Ibn-Tofail, K\'{e}nitra;$^{(c)}$Facult\'e des Sciences Semlalia, Universit\'e Cadi Ayyad, LPHEA-Marrakech;$^{(d)}$LPMR, Facult\'e des Sciences, Universit\'e Mohamed Premier, Oujda;$^{(e)}$Facult\'e des sciences, Universit\'e Mohammed V, Rabat;$^{(f)}$Institute of Applied Physics, Mohammed VI Polytechnic University, Ben Guerir; Morocco.\\
% ...
$^{46}$Physics Department, Southern Methodist University, Dallas TX; United States of America.\\
% ...
$^{50}$Fakult\"{a}t Physik , Technische Universit{\"a}t Dortmund, Dortmund; Germany.\\
% ...
$^{56}$II. Physikalisches Institut, Georg-August-Universit\"{a}t G\"ottingen, G\"ottingen; Germany.\\
% ...
$^{105}$CPPM, Aix-Marseille Universit\'e, CNRS/IN2P3, Marseille; France.\\
% ...
$^{120}$$^{(a)}$New York University Abu Dhabi, Abu Dhabi;$^{(b)}$United Arab Emirates University, Al Ain; United Arab Emirates.\\
% ...
$^{124}$Homer L. Dodge Department of Physics and Astronomy, University of Oklahoma, Norman OK; United States of America.\\
% ...
$^{158}$Raymond and Beverly Sackler School of Physics and Astronomy, Tel Aviv University, Tel Aviv; Israel.\\
% ...
$^{a}$ Also Affiliated with an institute covered by a cooperation agreement with CERN.\\
$^{b}$ Also at An-Najah National University, Nablus; Palestine.\\
$^{c}$ Also at Borough of Manhattan Community College, City University of New York, New York NY; United States of America.\\
$^{d}$ Also at Center for High Energy Physics, Peking University; China.\\
$^{e}$ Also at Center for Interdisciplinary Research and Innovation (CIRI-AUTH), Thessaloniki; Greece.\\
$^{f}$ Also at CERN, Geneva; Switzerland.\\
% ...

\end{flushleft}

% Created with Glance <Atlas.Glance@cern.ch>

%

%
%
\clearpage

\end{document}